\documentclass[prd,superscriptaddress,nofootinbib,floatfix,notitlepage]{revtex4-1}
\pdfoutput=1
\usepackage{amsmath,amssymb,epsfig,empheq}
\usepackage{hyperref}
\usepackage{natbib,ifthen}

\usepackage{wasysym}
\usepackage{mathrsfs}
\usepackage{simplewick}
\usepackage[lofdepth,lotdepth,caption=false]{subfig}
\usepackage{color}
\usepackage{bbold}

\begin{document}

% journals
\newcommand{\jcap}{JCAP}
\newcommand{\apjl}{APJL~}

% ms commands
\def\Btheo{{B_\delta^{\textrm{theo}}}}
\newcommand{\todo}[1]{{\color{blue}{TODO: #1}}} % write in main text
\newcommand{\mscomment}[1]{{\color{blue}{MS: #1}}} % write in main text
\newcommand{\zvcomment}[1]{{\color{blue}{ZV: #1}}} % write in main text
\newcommand{\lbar}[1]{\underline{l}_{#1}}
\newcommand{\drm}{\mathrm{d}}
\renewcommand{\d}{\mathrm{d}}
\renewcommand{\rm}[1]{\mathrm{#1}}
\newcommand{\gaensli}[1]{\lq #1\rq$ $}
\newcommand{\bartilde}[1]{\bar{\tilde #1}}
\newcommand{\barti}[1]{\bartilde{#1}}
\newcommand{\ti}{\tilde}
\newcommand{\oforder}[1]{\mathcal{O}(#1)}
\newcommand{\D}{\mathrm{D}}
\renewcommand{\(}{\left(}
\renewcommand{\)}{\right)}
\renewcommand{\[}{\left[}
\renewcommand{\]}{\right]}
\def\<{\left\langle}
\def\>{\right\rangle}
\newcommand{\mycaption}[1]{\caption{\footnotesize{#1}}}
\newcommand{\hattilde}[1]{\hat{\tilde #1}}
\newcommand{\mycite}[1]{[#1]}
\newcommand{\mnras}{Mon.\ Not.\ R.\ Astron.\ Soc.}
\newcommand{\apjs}{Astrophys.\ J.\ Supp.}

%%%%%% Vectors %%%%%%
\def\uk{{\bf \hat{k}}}
\def\un{{\bf \hat{n}}}
\def\ur{{\bf \hat{r}}}
\def\ux{{\bf \hat{x}}}
\def\bk{{\bf k}}
\def\bn{{\bf n}}
\def\br{{\bf r}}
\def\bx{{\bf x}}
\def\bK{{\bf K}}
\def\by{{\bf y}}
\def\bl{{\bf l}}
\def\bkp{{\bf k^\pr}}
\def\brp{{\bf r^\pr}}

% other commands
\newcommand{\fixme}[1]{{\textbf{Fixme: #1}}}
\newcommand{\detD}{{\det\!\cld}}
\newcommand{\clh}{\mathcal{H}}
\newcommand{\ud}{{\rm d}}
\renewcommand{\eprint}[1]{\href{http://arxiv.org/abs/#1}{#1}}
\newcommand{\adsurl}[1]{\href{#1}{ADS}}
\newcommand{\ISBN}[1]{\href{http://cosmologist.info/ISBN/#1}{ISBN: #1}}
\newcommand{\vort}{\varpi}
\newcommand\ba{\begin{eqnarray}}
\newcommand\ea{\end{eqnarray}}
\newcommand\be{\begin{equation}}
\newcommand\ee{\end{equation}}
\newcommand\lagrange{{\cal L}}
\newcommand\cll{{\cal L}}
\newcommand\cln{{\cal N}}
\newcommand\clx{{\cal X}}
\newcommand\clz{{\cal Z}}
\newcommand\clv{{\cal V}}
\newcommand\cld{{\cal D}}
\newcommand\clt{{\cal T}}

\newcommand\clo{{\cal O}}
\newcommand{\cla}{{\cal A}}
\newcommand{\clp}{{\cal P}}
\newcommand{\clr}{{\cal R}}
\newcommand{\uD}{{\mathrm{D}}}
\newcommand{\calE}{{\cal E}}
\newcommand{\calB}{{\cal B}}
\newcommand{\curl}{\,\mbox{curl}\,}
\newcommand\del{\nabla}
\newcommand\Tr{{\rm Tr}}
\newcommand\half{{\frac{1}{2}}}
\newcommand\fourth{{1\over 8}}
\newcommand\bibi{\bibitem}
\newcommand{\kf}{\beta}
\newcommand{\kfprod}{\alpha}
\newcommand\calS{{\cal S}}
\renewcommand\H{{\cal H}}
\newcommand\K{{\rm K}}
\newcommand\mK{{\rm mK}}
\newcommand\synch{\text{syn}}
\newcommand\opacity{\tau_c^{-1}}

\newcommand{\Psil}{\Psi_l}
\newcommand{\bsigma}{{\bar{\sigma}}}
\newcommand{\bI}{\bar{I}}
\newcommand{\bq}{\bar{q}}
\newcommand{\bv}{\bar{v}}
\renewcommand\P{{\cal P}}
\newcommand{\numfrac}[2]{{\textstyle \frac{#1}{#2}}}

\newcommand{\la}{\langle}
\newcommand{\ra}{\rangle}
\newcommand{\lla}{\left\langle}
\newcommand{\rra}{\right\rangle}

\newcommand{\vnabla}{\ensuremath{\boldsymbol\nabla}}

\newcommand{\Omtot}{\Omega_{\mathrm{tot}}}
\newcommand\xx{\mbox{\boldmath $x$}}
\newcommand{\phpr} {\phi'}
\newcommand{\gam}{\gamma_{ij}}
\newcommand{\sqgam}{\sqrt{\gamma}}
\newcommand{\delk}{\Delta+3{\K}}
\newcommand{\dph}{\delta\phi}
\newcommand{\om} {\Omega}
\newcommand{\dom}{\delta^{(3)}\left(\Omega\right)}
\newcommand{\rar}{\rightarrow}
\newcommand{\Rar}{\Rightarrow}
\newcommand\gsim{ \lower .75ex \hbox{$\sim$} \llap{\raise .27ex \hbox{$>$}} }
\newcommand\lsim{ \lower .75ex \hbox{$\sim$} \llap{\raise .27ex \hbox{$<$}} }
\newcommand\bigdot[1] {\stackrel{\mbox{{\huge .}}}{#1}}
\newcommand\bigddot[1] {\stackrel{\mbox{{\huge ..}}}{#1}}
\newcommand{\Mpc}{\text{Mpc}}
\newcommand{\Al}{{A_l}}
\newcommand{\Bl}{{B_l}}
\newcommand{\eAl}{e^\Al}
\newcommand{\ix}{{(i)}}
\newcommand{\ixp}{{(i+1)}}
\renewcommand{\k}{\beta}
% Derivatives
\newcommand{\HD}{\mathrm{D}}

\newcommand{\nonflat}[1]{#1}
\newcommand{\Cgl}{C_{\text{gl}}}
\newcommand{\Cgltwo}{C_{\text{gl},2}}
\newcommand{\He}{{\rm{He}}}
\newcommand{\Mhz}{{\rm MHz}}
\newcommand{\vx}{{\mathbf{x}}}
\newcommand{\ve}{{\mathbf{e}}}
\newcommand{\vv}{{\mathbf{v}}}
\newcommand{\vk}{{\mathbf{k}}}
\renewcommand{\vr}{{\mathbf{r}}}
\newcommand{\vn}{{\mathbf{n}}}
\newcommand{\vPsi}{{\mathbf{\Psi}}}
\newcommand{\vs}{{\mathbf{s}}}
\newcommand{\vH}{{\mathbf{H}}}
\newcommand{\theo}{\mathrm{th}}
\newcommand{\sgn}{\mathrm{sgn}}

\newcommand{\vnhat}{{\hat{\mathbf{n}}}}
\newcommand{\vkhat}{{\hat{\mathbf{k}}}}
\newcommand{\taueps}{{\tau_\epsilon}}

\newcommand{\vgrad}{{\mathbf{\nabla}}}
\newcommand{\fbarln}{\bar{f}_{,\ln\epsilon}(\epsilon)}

% marcel commands
\newcommand{\secref}[1]{Section \ref{se:#1}}
\newcommand{\expt}{\mathrm{expt}}
\newcommand{\eq}[1]{(\ref{eq:#1})} 
\newcommand{\eqq}[1]{Eq.~(\ref{eq:#1})} 
\newcommand{\fig}[1]{Fig.~\ref{fig:#1}} 
\renewcommand{\to}{\rightarrow}
\renewcommand{\(}{\left(}
\renewcommand{\)}{\right)}
\renewcommand{\[}{\left[}
\renewcommand{\]}{\right]}
\renewcommand{\vec}[1]{\mathbf{#1}}
\newcommand{\vy}{\vec{y}}
\newcommand{\vz}{\vec{z}}
\newcommand{\vq}{\vec{q}}
\newcommand{\vp}{\vec{p}}
\newcommand{\va}{\vec{a}}
\newcommand{\vb}{\vec{b}}
\newcommand{\VPsi}{\vec{\Psi}}
\newcommand{\vecv}{\vec{v}}
\newcommand{\vl}{\vec{l}}
\newcommand{\VL}{\vec{L}}
\newcommand{\dl}{\d^2\vl}
\newcommand{\valpha}{\vec{\alpha}}
\renewcommand{\L}{\mathscr{L}}

\newcommand{\abs}[1]{\lvert #1\rvert}
%\renewcommand{\bf}{\mathbf}

%% marcel commands
\newcommand{\ul}{\underline{l}}
\newcommand{\lin}{\mathrm{lin}}

%% zvonimir commands
\newcommand*{\df}  {\delta}
\newcommand*{\tf}  {\theta}
\renewcommand{\vec}{\textbf}
\newcommand*{\non}  {\nonumber}

\def\pvm#1{{\color{blue}[PM: {\it #1}] }}

%%%%%%%%%%%%%%%%%%%%%%%%%%%%%%%%%%%%%%%%%%%%%%%%

\thispagestyle{empty}

\title{Fast Large Scale Structure Perturbation Theory using 1D FFTs}

\author{Marcel Schmittfull}
\affiliation{Berkeley Center for Cosmological Physics, Department of
  Physics and Lawrence Berkeley
  National Laboratory, University of California, Berkeley, CA 94720, USA}

\author{Zvonimir Vlah}
\affiliation{Stanford Institute for Theoretical Physics and Department of Physics, Stanford University, Stanford, CA 94306, USA}
\affiliation{Kavli Institute for Particle Astrophysics and Cosmology, SLAC and Stanford University, Menlo Park, CA 94025, USA}

\author{Patrick McDonald}
\affiliation{Lawrence Berkeley National Laboratory, One Cyclotron Road, Berkeley, CA 94720, USA}

\date{\today}

\begin{abstract}

The usual fluid equations describing the large-scale evolution of mass density 
in the universe can be written as local in the density, velocity divergence, 
and velocity potential fields.
As a result, the perturbative expansion in small
density fluctuations, usually written in terms of convolutions in Fourier 
space, can be written as a series of products of these fields evaluated at 
the same location in configuration space.
Based on this, we establish a new method to numerically 
evaluate the 1-loop power spectrum (i.e., Fourier transform of the 2-point correlation 
function) with one-dimensional Fast Fourier Transforms.  This is exact and a 
few orders of magnitude faster than previously used numerical approaches. 
Numerical results of the new method are in excellent agreement 
with the standard quadrature integration method. 
This fast model evaluation can in principle be extended to higher loop order where 
existing codes become painfully slow.
Our approach follows by writing higher order corrections to the 2-point 
correlation function as, e.g., the correlation between two second-order 
fields or the correlation between a linear and a third-order field. 
These are then decomposed into products of correlations of linear fields and 
derivatives of linear fields.
The method can also be viewed as evaluating three-dimensional Fourier space 
convolutions using products in configuration space, which may also be useful
in other contexts where similar integrals appear. 

\end{abstract}

\maketitle

\section{Introduction}

Observations of the large-scale structure (LSS) of the universe play an important role in cosmology.  In the near future, a plethora of LSS surveys including e.g.~DES \cite{DESwhitepaper}, eBOSS \cite{eBOSSDawson}, DESI \cite{DESIwhitepaper}, Euclid \cite{EuclidWhitePaper}, WFIRST \cite{WFIRST1503}, LSST \cite{LSSTDESC}, and possibly SPHEREx \cite{spherex1412} are planning to improve our understanding of the nature of dark energy, the origin of the universe, and neutrino 
properties.  Their science return strongly depends on the precision of 
models for summary statistics of the LSS, especially in the nonlinear regime where many modes are observed but modeling is rather challenging.  Commonly used perturbative models such as Eulerian standard perturbation theory (SPT) (e.g.~\cite{Goroff:1986ep, Jain:1993jh, 1996ApJS..105...37S, 1996ApJ...473..620S, Blas:2013bpa}) or Lagrangian perturbation theory (LPT) (e.g.~\cite{Zeldovich:1969sb, Bouchet:1994xp, Matsubara:2007wj}) fail to describe summary statistics on nonlinear scales if they are truncated at the lowest perturbative order; see e.g.~\cite{bernardeauReview} for a review and \cite{CarlsonWhitePadmanabhan0905} for simulation comparison.  The range of validity of these models extends significantly into the mildly nonlinear regime when including higher-order corrections.  Modern cosmological LSS analyses often include such corrections to include smaller scales in the analysis and thus improve cosmological constraints.

The modeling improvement from higher order corrections comes however at the price of higher numerical complexity and computational cost. 
Since LSS perturbation theory is used on a daily basis by a large community of researchers working on LSS, it is useful to simplify and speed up the numerical evaluation of these model corrections.
In addition to the practical value of simplifying and speeding up higher order 
corrections, the expressions we derive also have a useful physical 
interpretation.
For example, power spectrum corrections at fourth order in the linear density involve products of 2-point correlation functions of linear fields and derivatives of linear fields.  
Due to statistical isotropy, these 2-point correlation functions depend only on the separation length between the two fields, and not on the orientation of the separation vector. 
They can be interpreted as the coefficients that arise when decomposing the 
2-point correlation function between field derivatives into irreducible 
components of the separation vector. 

This approach is also useful 
to simplify 2-loop or higher-order loop corrections to the power spectrum, where Fourier space integrals naively become rather high dimensional and numerical evaluation can take several hours just for a single cosmology.  
For 3-point statistics, the loop integrals have to be performed for a large number of wavevector or separation triplets, again leading to high computational cost; and the problem only becomes worse for higher-order $N$-point functions that may be needed e.g.~for covariances.

Adding to the motivation above, a concrete example where simpler and faster higher-order LSS models are beneficial is the exploration of cosmological parameter space with sampling methods such as Markov Monte Carlo (MCMC) chains to interpret LSS observations. 
     In practice, data analyses may go through various iterations of updated catalogs, data splits, fitting procedures, prior or likelihood choices, and model extensions beyond the standard $\Lambda$CDM model. The human and computer time required for these steps can be reduced if the total time needed for MCMC chains is reduced by faster model evaluations. In the CMB community, examples for similar modeling speedups include CMBFAST \cite{cmbfast} and CAMB \cite{camb}, or 
 approximate methods like Pico \cite{pico}. These tools have been crucial for 
cosmology from CMB observations in the past decade.

Motivated by these points, the main goal of this paper is to provide a fast algorithm to evaluate nonlinear loop corrections for models of the LSS of the universe.  The main idea of our approach can be illustrated for the 1-loop power spectrum correction in SPT arising from correlations of two second-order densities. This 1-loop integral has contributions of the form
\begin{align}
  \label{eq:P22intro}
  P_{22}(k) \sim \int\frac{\d^3\vq}{(2\pi)^3} f(q)g(|\vk-\vq|)P_\lin(q)P_\lin(|\vk-\vq|),
\end{align}
where $P_\lin$ is the linear power spectrum and $f$ and $g$ are general functions whose exact form is not important.\footnote{The exact form of $f$ and $g$ is determined by the equations of motion for the dark matter fluid. The full $P_{22}$ integrand has additional angular dependence that we omit here for simplicity; see \secref{Fast22} for the full expression. }  The integral in \eqq{P22intro} is a three-dimensional convolution in Fourier space. 
It can be evaluated efficiently by Fourier transforming $fP_\mathrm{lin}$ and $gP_\mathrm{lin}$ to configuration space, multiplying them there, and Fourier transforming the result back to $k$ space.  Due to isotropy of the problem, the angular parts of the three-dimensional Fourier transforms are independent of $P_\lin$ and can be performed analytically independent of cosmology. We are then only left with the radial parts of three-dimensional Fourier transforms, which reduce to one-dimensional Hankel transforms:
\begin{eqnarray}
  P_{22}(k) \sim   \int_0^\infty \d r\,r^2 j_0(kr)
\left[\int_0^\infty \d q\, q^2\, j_0(qr) f(q)P_\mathrm{lin}(q) \right]
  \label{eq:P22_intro_fast}
\left[\int_0^\infty \d p\, p^2 \,j_0(pr) g(p)P_\mathrm{lin}(p) \right].
\end{eqnarray}
Each of the two square brackets is a one-dimensional Hankel transform of a 
filtered linear power spectrum, whose result are two functions of $r$. The 
integral of their product over $r$ is then again a one-dimensional Hankel 
transform.
Each one-dimensional Hankel transform can be evaluated efficiently with a one-dimensional Fast Fourier Transform (FFT) by using FFTLog \cite{hamiltonfftlog} or similar numerical packages.  
The algorithm to evaluate the  part of $P_{22}(k)$ outlined above thus scales as $\mathcal{O}(3 N\log N)$, where $N$ is the number of grid points where $P_\lin$ is evaluated.   These $\sim 3N\log N$ operations return the integral evaluated at all $k$ bins at once.  This should be contrasted with the standard procedure to evaluate this type of integral, where a two-dimensional integral over radius and angle is needed for every $k$ bin, which scales naively like $\mathcal{O}(N^3)$.  Although tuning the evaluation points can somewhat improve this, we generally expect the one-dimensional FFT algorithm to be at least a few hundred times faster.  
Since no approximations go into \eqq{P22_intro_fast}, the algorithm is also exact.  
We show in the rest of the paper how this approach can be applied to all nonlinear power spectrum corrections that arise at fourth order in the linear density in SPT.  
The efficient performance of FFTLog has also recently been used in similar contexts, where the mapping of Lagrangian displacement cumulants to 
the matter power spectrum was also expressed as an expansion in Hankel transforms \cite{zvonimir1410, Vlah:2015sea}.

As discussed in Appendix \ref{se:configspaceevolution}, the possibility of 
this form of evaluation is not an accident, but a consequence of the fact that
the underlying evolution equations for the fields take the form of local 
products, which can be evaluated by trivial multiplication in configuration 
space, combined with simple derivative operators that can be applied
efficiently in Fourier space. The complex-looking wave-vector integrals over
kernels in the standard PT 
formulation are nothing more than the Fourier space representation of these
local field products and derivatives. 
There is a finite number of such integrals that need to be
evaluated because the evolution equations only involve a small set of
derivative operators acting on the fields.

The paper is organized as follows. We start with a brief introduction to Eulerian Standard Perturbation Theory (SPT).  \secref{Fast22} shows how to evaluate correlations between two second-order densities in a fast way, describing one example term in detail, followed by a more general treatment of the $P_{22}$ correction in SPT.  \secref{Fast13} shows similar results for the $P_{13}$ correlation between linear and third-order densities. Numerical results are presented in \secref{numerics}. We conclude in \secref{conclusions}. 
Appendices provide evolution equations for the dark matter fluid motivating our
configuration space approach, fast expressions for some displacement statistics
appearing in Lagrangian perturbation theory, and useful mathematical 
identities.

\subsection*{Conventions and notation}
Throughout the paper,  $\vx$ and $\vx'$ denote configuration space positions, and $\vr$ denotes configuration space separations. The variables $\vk$, $\vq$ and $\vp$ are used for Fourier space wavevectors. We  use Fourier conventions
\begin{align}
  \label{eq:47}
  f(\vk) = \int \d^3\vr\, e^{i\vk\cdot\vr}\,f(\vr),
\qquad 
f(\vr) = \int_\vk e^{-i\vk\cdot\vr}\,f(\vk)
\end{align}
with the shorthand notation
\begin{align}
  \label{eq:49}
  \int_\vk \equiv \int\frac{\d^3\vk}{(2\pi)^3}.
\end{align}
Derivative operators act only on the single variable to their immediate right.  
Vectors with hats are unit vectors, e.g.~$\hat\vk=\vk/k$.

\section{Eulerian Standard Perturbation Theory (SPT)}

In Eulerian Standard Perturbation Theory (SPT), the nonlinear dark matter overdensity is expanded as
\begin{align}
  \label{eq:deltaNLx}
  \delta(\vx) = \sum_{n=1}^{\infty}\delta^{(n)}(\vx),
\end{align}
where $\delta^{(1)}=\delta_0$ is the linear density and $\delta^{(n)}=\mathcal{O}(\delta_0^n)$. The second order part has nonlinear growth, shift and tidal contributions:
\begin{align}
  \label{eq:delta2x}
\delta^{(2)}(\vx) = \frac{17}{21}\delta_0^2(\vx)+ \vPsi(\vx)\cdot \vnabla\delta_0(\vx)+\frac{4}{21}s^2(\vx).
\end{align}
The shift term contains the linear displacement field
\begin{align}
  \label{eq:46}
  \vPsi(\vx) = \frac{\vnabla}{\nabla^2}\delta_0(\vx),
\end{align}
and the tidal term is given by
\begin{eqnarray}
  s^2(\vx) = \frac{3}{2}
\left(\frac{\nabla_i\nabla_j}{\nabla^2}-\frac{1}{3}\delta_{ij}^\mathrm{(K)}\right)
\delta_0(\vx)
\left(\frac{\nabla_i\nabla_j}{\nabla^2}-\frac{1}{3}\delta_{ij}^\mathrm{(K)}\right)\delta_0(\vx).
\end{eqnarray}
The configuration space products in \eqq{delta2x} are commonly written as Fourier space convolutions involving the $F_2$ kernel,\footnote{In our conventions, $\vPsi(\vk) = \frac{i\vk}{k^2}\delta_0(\vk)$, $[\vnabla\delta](\vk)=-i\vk\delta(\vk)$ and $s^2(\vx)=\frac{3}{2}s_{ij}(\vx)s_{ij}(\vx)$ with $s_{ij}(\vk)=[k_ik_j/k^2-\delta^\mathrm{(K)}_{ij}/3]\delta_0(\vk)$.  Then $\delta(\vk) = \delta_0(\vk)+\int_\vq F_2(\vq,\vk-\vq)\delta_0(\vq)\delta_0(\vk-\vq)+\mathcal{O}(\delta_0^3)$, where $F_2(\vq,\vk-\vq)=\frac{17}{21} + \frac{1}{2}\left(\frac{q}{|\vk-\vq|}+\frac{|\vk-\vq|}{q}\right)\mu+\frac{4}{21}\frac{3}{2}(\mu^2-\frac{1}{3})$ with $\mu\equiv\hat\vq\cdot(\widehat{\vk-\vq})$.} but it is simpler for now to continue in configuration space. 

The nonlinear 2-point correlation function up to fourth order in the linear density is
\begin{eqnarray}
\la\delta(\vx)\delta(\vx+\vr)\ra 
=
 \la\delta_0(\vx)\delta_0(\vx+\vr)\ra 
+\la\delta^{(2)}(\vx)\delta^{(2)}(\vx+\vr)\ra 
+2\la\delta^{(1)}(\vx)\delta^{(3)}(\vx+\vr)\ra + \mathcal{O}(\delta_0^6),
\end{eqnarray}
where we assumed Gaussian linear density $\delta_0$.
We write this as
\begin{align}
  \label{eq:48}
  \xi(r) = \xi_\lin(r) + \xi_{(22)}(r) + \xi_{(13)}(r) + 
\mathcal{O}(\delta_0^6).
\end{align}
$\xi_\lin$ is the tree-level correlation function arising from the correlation between two linear densities separated by $r$. $\xi_{(22)}$ is the 1-loop correction arising from the correlation between two second order densities, and $\xi_{(13)}$ is the 1-loop correction from the correlation between linear and third order density.

\section{Fast evaluation of 2-2 terms}
\label{se:Fast22}

\subsection{An example}
\label{se:22Example}

Before computing fast expressions for the full 2-2 correlation function and power spectrum, we consider a simple example term to illustrate the general idea of our approach:\footnote{Related calculations in the literature, e.g.~\cite{sherwinZaldarriaga,Mccollough1202}, will be discussed in \secref{P22literature}.}
\begin{align}
  \label{eq:xi22demostep1}
\xi_{(22)}(r)&= \frac{17}{21}\,\big\la \vPsi(\vx)\cdot\vnabla\delta_0(\vx)\, \delta_0(\vx+\vr)\delta_0(\vx+\vr)\big\ra+\cdots\\
\label{eq:xi22demostep2}
&= \frac{34}{21}\, 
\big\la \vPsi(\vx)\delta_0(\vx+\vr) \big\ra\,\cdot\,
\big\la\vnabla\delta_0(\vx)\delta_0(\vx+\vr)\big\ra
+\cdots.
\end{align}
This contribution involves the correlation between linear displacement and linear density, and the correlation between linear density gradient and linear density. We compute these correlation functions in a way that is particularly suited for numerical evaluation by Fourier transforming the linear fields:
\begin{eqnarray}
  \big\la \vPsi(\vx)\delta_0(\vx+\vr) \big\ra &=&
\int_{\vp,\vq}e^{-i\vp\cdot\vx-i\vq\cdot(\vx+\vr)} \left\la  \frac{i\vp}{p^2} \delta_0(\vp) \delta_0(\vq)\right\ra\non\\
\label{eq:demoCorrelPsiDeltaStep2}
&=&
\int_{\vq}e^{-i\vq\cdot\vr}\,\frac{-i\vq}{q^2}\,P_\lin(q)
\\
\label{eq:demoCorrelPsiDeltaStep3}
&=&-\hat\vr\,i\int_{\vq}e^{-i\vq\cdot\vr}\,q^{-1}\,\hat\vq\cdot\hat\vr\,P_\lin(q)  \\
\label{eq:demoCorrelPsiDeltaStep4}
&=& -\hat\vr\,\xi^1_{-1}(r).
\end{eqnarray}
\eqq{demoCorrelPsiDeltaStep3} follows by writing \eqq{demoCorrelPsiDeltaStep2} in terms of irreducible components of $\hat\vr$, similarly to calculations in \cite{zvonimir1410}.
In \eqq{demoCorrelPsiDeltaStep4} we defined the generalized linear correlation function\footnote{This is defined for arbitrary integer $n$ and non-negative integer $l$. Note $\xi^0_0(r)=\xi_\lin(r)$, and see \secref{xiinterpretation} for interpretation. The inverse relation of \eqq{xidef} is
\begin{align}
  \label{eq:34}
  k^nP_\lin(k) = 4\pi\int_0^\infty \d r\,r^2 j_l(kr)\xi^l_n(r).
\end{align}
} 
\begin{eqnarray}
  \xi^l_n(r) &\equiv& 
  \label{eq:xidef3D}
 i^l \int_\vq e^{-i\vq\cdot\vr} \,q^n\, \mathsf{P}_l(\hat\vq\cdot\hat\vr)\,P_\lin(q)\\
  \label{eq:xidef}
&=& \int_0^\infty \frac{\d q}{2\pi^2}\,q^{2+n}\,j_l(qr)\,P_\lin(q).
\end{eqnarray}
The angular integration follows from \eqq{IntegrateExpLegendreOverAngle}. $\mathsf{P}_l$ is the $l$th-order Legendre polynomial and $j_l$ is the spherical Bessel function. Crucially, \eqq{xidef} is a 1D Hankel transform of the linear power spectrum.  This can be evaluated numerically with a 1D Fast Fourier Transform (FFT) using FFTLog \cite{hamiltonfftlog}.  Similarly, the other correlation function entering \eqq{xi22demostep2} becomes
\begin{align}
  \label{eq:grad_delta_delta}
  \la \vnabla\delta_0(\vx)\,\delta_0(\vx+\vr) \ra
= \hat\vr\,\xi^1_1(r).
\end{align}
Then, the 2-2 correlation written out in \eqq{xi22demostep2} becomes
\begin{align}
  \label{eq:55}
  \xi_{(22)}(r) = -\frac{34}{21}\xi_{-1}^1(r)\xi^1_1(r)+\cdots.
\end{align}

The corresponding power spectrum contribution is 
\begin{eqnarray}
  P_{22}(k) &=& \int\d^3\vr\,e^{i\vk\cdot\vr}\,\xi_{(22)}(r) \\
  \label{eq:P22fromxi22}
&=& 4\pi\int_0^\infty \d r\,r^2\,j_0(kr)\,\xi_{(22)}(r).
\end{eqnarray}
This is again a 1D Hankel transform that can be evaluated as a 1D FFT with FFTLog.  The particular contribution to $P_{22}(k)$ from \eqq{xi22demostep2} can thus be computed at all desired output values $k$ with only three 1D FFTs. This is orders of magnitude faster than conventional integration methods for $P_{22}$, which perform 2D integrals for every value of $k$.

\subsection{Full 2-2 correlation function and power spectrum in SPT}

The example calculation above only considered a single contribution to the 2-2 correlation function. 
Other contributions arise from all possible correlations between growth, shift and tidal term in \eqq{delta2x}.
To calculate these correlations, we only need to consider correlations between derivatives of the linear density and the linear density of the form
\begin{align}
  \label{eq:correlfor22}
  \left\la \left[
\nabla^n\frac{\nabla_{i_1}\cdots\nabla_{i_L}}{\nabla^L}\delta_0(\vx) \right]
\delta_0(\vx+\vr)
\right\ra.
\end{align}
This is sufficiently general because derivative operators can be moved from one $\delta_0$ to the other without changing the resulting expectation value (up to a sign).\footnote{For example, $\la\vPsi(\vx)\nabla\delta_0(\vx+\vr)\ra=-\la[\nabla\vPsi(\vx)]\delta_0(\vx+\vr)\ra$. This follows by rewriting the expectation value in Fourier space, or from integration by parts when replacing the ensemble average by an average over positions $\vx$.}
We assume that $n-L$ is even, so that $\nabla^{n-L}=(\nabla^2)^{(n-L)/2}$ corresponds to the Laplace operator raised to some power.
Generalizing \eqq{demoCorrelPsiDeltaStep3}, the contraction of the correlation function \eq{correlfor22} with $\hat\vr_{i_1}\cdots\hat\vr_{i_L}$ is given by a sum over  general correlation functions $\xi^l_n$'s: 
\begin{align}
  \label{eq:xiPhysics}
\sum_{i_1\cdots i_L}
  \left\la \left[
\nabla^n\frac{\nabla_{i_1}\cdots\nabla_{i_L}}{\nabla^L}\delta_0(\vx) \right]
\delta_0(\vx+\vr)
\right\ra  \hat\vr_{i_1}\cdots\hat\vr_{i_L}
=
-\sum_{l=0}^L (2l+1) \,\alpha_{Ll}\,\xi^l_{n}(r).
\end{align}
The correlation between the derivative field $\nabla^n \hat\nabla^L\delta_0$ and the linear density $\delta_0$ thus generates $\xi^l_n$'s with $l\le L$.
The $\alpha$ coefficients arise when changing bases to Legendre polynomials; see \eqq{ScalprodInPls} below.

Applying \eqq{xiPhysics} to all 2-2 correlations gives the following simple expression for the full 2-2 correlation function in Eulerian SPT:
\begin{empheq}[box=\fbox]{align}
  \xi_{(22)}(r) &= \frac{1219}{735}[\xi_0^0(r)]^2
+\frac{1}{3}\xi_{-2}^0(r)\xi_{2}^0(r)
-\frac{124}{35}\xi_{-1}^1(r)\xi_{1}^1(r)
+\frac{1342}{1029}[\xi_0^2(r)]^2
+\frac{2}{3}\xi_{-2}^2(r)\xi_{2}^2(r)
\non\\
\label{eq:xi22fast}
& \quad\,
-\frac{16}{35} \xi_{-1}^3(r)\xi_{1}^3(r)
+ \frac{64}{1715}[\xi_0^4(r)]^2.
\end{empheq}
Using Eqs.~\eq{P22fromxi22} and \eq{xi22fast}, the calculation of $P_{22}$ is thus reduced to computing twelve 1D Hankel transforms, which can be evaluated efficiently and robustly in terms of 1D FFTs with FFTLog \cite{hamiltonfftlog}.
The computational cost to compute $P_{22}(k)$ at all wavenumbers $k$ then scales as 
$\mathcal{O}(12\,N\log N)$ if the linear power spectrum is sampled at $N$ 
points. This should be significantly faster than the standard procedure that 
calculates $P_{22}(k)$ with a 2D integral for every $k$, naively using 
$\mathcal{O}(N^3)$ operations. Although there may be substantial differences
in prefactors in either direction, 
we naively expect a speedup 
factor of $\mathcal{O}(1000)$ for $N=100$ evaluation points.

\subsection{Interpretation of generalized correlation functions \texorpdfstring{$\xi^l_n$}{xi^l_n}}
\label{se:xiinterpretation}

The $\xi^l_n$ correlation functions defined in \eqq{xidef} can be interpreted by noting that they satisfy
\begin{align}
  \label{eq:xiln_interpretation}
  \xi^l_n(r)=(-i)^{l+n} (-1)^L\, \sum_{i_1\cdots i_L}T^{(l)}_{i_1\cdots i_L}(\hat\vr)\,
\left\la \left[
\nabla^n\frac{\nabla_{i_1}\cdots\nabla_{i_L}}{\nabla^L}\delta_0(\vx) \right]
\delta_0(\vx+\vr)
\right\ra,
\end{align}
where $T^{(l)}_{i_1\cdots i_L}(\hat\vr)$ are irreducible tensors defined by
$T_{i}^{(1)}(\hat\vr)=\hat\vr_i$,
$T_{ij}^{(0)}(\hat\vr)=\delta_{ij}$, 
$T_{ij}^{(2)}(\hat\vr)=\tfrac{3}{2}\left( \hat\vr_i\hat\vr_j-\frac{\delta_{ij}}{3}\right)$ etc., and normalized so that
\begin{align}
  \label{eq:2}
\sum_{i_1\cdots i_L}  \hat\vq_{i_1}\cdots\hat\vq_{i_L}\,  T^{(l)}_{i_1\cdots i_L}(\hat\vr) = \mathsf{P}_l(\hat\vq\cdot\hat\vr).
\end{align}
\eqq{xiln_interpretation} then follows by writing the correlation function \eq{correlfor22} in Fourier space and contracting it with these irreducible tensors.
The inverse relation of \eqq{xiln_interpretation} for suitably normalized tensors $\tilde T(\hat\vr)\propto T(\hat\vr)$ is
\begin{align}
  \label{eq:correlFcnDecomp}
  \left\la \left[
\nabla^n\frac{\nabla_{i_1}\cdots\nabla_{i_L}}{\nabla^L}\delta_0(\vx) \right]
\delta_0(\vx+\vr)
\right\ra 
= \sum_{l=0}^L \xi^l_n(r) \,\tilde T^{(l)}_{i_1\cdots i_L}(\hat\vr).
\end{align}
This follows by contracting both sides with $T(\hat\vr)$, and generalizes Eqs.~\eq{demoCorrelPsiDeltaStep4} and \eq{grad_delta_delta}.

\eqq{correlFcnDecomp} shows that the linear correlation functions $\xi^l_n(r)$ can be interpreted as the expansion coefficients that appear when decomposing the 2-point correlation function between derivatives of the linear density into irreducible components of the separation vector $\vr$. 
Due to statistical isotropy, these coefficients $\xi^l_n(r)$ only depend on the separation length $r$ but not on the orientation of the separation.

\subsection{Alternative Fourier space derivation of fast $P_{22}$}

In the last section we derived the fast expression  \eq{xi22fast} for the 2-2 correlation function in SPT by working mostly in configuration space. In this section we present an alternative derivation of the same result by working entirely in Fourier space.

The 2-2 SPT power spectrum in Fourier space is
\begin{align}
    \label{eq:P22_def}
  P_{22}(k) = 
2\int_{\vq}[F_2(\vq,\vk-\vq)]^2 P_\lin(q)P_\lin(|\vk-\vq|).
\end{align}
Expanding the square of the symmetrized second-order $F_2$ kernel in Legendre polynomials gives
\begin{align}
  2[F_2(\vq,\vp)]^2 = &
  \left[\frac{1219}{735}+\frac{1}{6}\left(\frac{q^2}{p^2} +
  \frac{p^2}{q^2}\right)\right]\mathsf{P}_0(\mu)
 +
  \frac{62}{35}\left(\frac{q}{p}+\frac{p}{q}\right)\mathsf{P}_1(\mu)
 + \left[\frac{1342}{1029} +
  \frac{1}{3}\left(\frac{q^2}{p^2}+\frac{p^2}{q^2}\right)\right]\mathsf{P}_2(\mu)\non\\
&+
  \frac{8}{35}\left(\frac{q}{p}+\frac{p}{q}\right)\mathsf{P}_3(\mu)
 +\frac{64}{1715}\mathsf{P}_4(\mu),
\label{eq:F2sqInLegendre}
\end{align}
where we introduced $\vp\equiv \vk-\vq$ and $\mu\equiv \hat{\vq}\cdot\hat{\vp}$. 
The fast $P_{22}$ expression \eq{xi22fast} then follows from the general identity 
\begin{align}
\label{eq:P22FormFinal}
\int_{\vq}q^{n_1}\,|\vk-\vq|^{n_2}\,\mathsf{P}_l(\hat\vq\cdot(\widehat{\vk-\vq}))
\,P_\lin(q)\,P_\lin(|\vk-\vq|)
= (-1)^l\, 4\pi \int_0^\infty \d r\,r^2\, j_0(kr)\,\xi^l_{n_1}(r)\, \xi^l_{n_2}(r).
\end{align}
This expresses the Fourier space convolution on the left hand side in terms of the Fourier transform of a configuration space product on the right hand side.\footnote{One way to obtain \eqq{P22FormFinal} is to 
introduce a Dirac delta to impose $\vp=\vk-\vq$, expand this Dirac delta in plane waves \eq{DiracDeltaInPlaneWaves}, and integrate over $\hat\vp$, $\hat\vq$ and $\hat\vr$ using \eqq{IntegrateExpLegendreOverAngle}, which is based on the addition theorem for spherical harmonics.
} 

The Legendre decomposition  \eq{F2sqInLegendre} was obtained by  converting between powers of wavevector scalar products and Legendre polynomials using
\begin{eqnarray}
  \label{eq:ScalprodInPls}
  (\hat\vx\cdot\hat\vy)^l &=& \sum_{l'=0}^l (2l'+1)\alpha_{ll'}  \mathsf{P}_{l'}(\hat\vx\cdot\hat\vy)\\
\label{eq:ScalprodInYlms}
&=&
4\pi\sum_{l'=0}^l\sum_{m'=-l}^l
\alpha_{ll'} Y_{l'm'}(\hat\vx) Y^*_{l'm'}(\hat\vy),
\end{eqnarray}
The crucial property of \eqq{ScalprodInYlms} is that the right hand side is separable, which can be exploited to separate integrals.
 The coefficients $\alpha_{ll'}$ describing the basis change between monomials and Legendre polynomials are given by \cite{zvonimir1410}
\begin{align}
  \label{eq:alphas}
\alpha_{ll'} = \frac{1}{2} \int_{-1}^1\d\mu \,\mu^l\, \mathsf{P}_{l'}(\mu)
=
  \begin{cases}
\frac{l!}{2^{(l-l')/2} \left[(l-l')/2\right]!\,
(l+l'+1)!!}, & \text{if $l\ge l'$ \& $l$ and $l'$ both even or odd}  ,\\
0, & \text{otherwise},
  \end{cases}
\end{align}
satisfying $\alpha_{ll'}=0$ if $l'>l$ or if $l+l'$ is odd (e.g.~$\alpha_{00}=1$, $\alpha_{10}=0$, $\alpha_{11}=1/3$, $\alpha_{20}=1/3$, $\alpha_{21}=0$, $\alpha_{22}=2/15$).
The general identity \eq{P22FormFinal} can then also be rewritten as
\begin{align}
  \label{eq:P22FormFinal2}
\int_{\vq}q^{n_1}|\vk-\vq|^{n_2} \left[\hat\vq\cdot(\widehat{\vk-\vq})\right]^{l} \,
P_\lin(q)P_\lin(|\vk-\vq|)
=
4\pi\int_0^{\infty}\d r\,r^2
j_0(kr)
\sum_{l'=0}^{l} (-1)^{l'}(2l'+1)\alpha_{ll'} \, \xi^{l'}_{n_1}(r)\,\xi^{l'}_{n_2}(r).
\end{align}
 As an example, we use this in Appendix~\ref{se:QnRnIntegrals}  to derive fast expressions for displacement statistics $Q_n$ relevant for Lagrangian Perturbation Theory.

The approach above can of course be applied to any integrals that have the general 3D convolution form of \eqq{P22FormFinal} or \eqq{P22FormFinal2}, which may appear in other contexts. The integrands can be slightly generalized to any general product-separable function of the form $f_1(q)f_2(p)$, in which case the factor functions $f_1$ and $f_2$ will enter \eqq{xidef} as 1D filters.
The Fourier transform approach can also be applied to lower- or higher-dimensional convolution integrals (see e.g.~upcoming CMB lensing work \cite{BoehmN32} where some 2D convolution integrals will be evaluated with FFTs).

\subsection{Related calculations in the literature}
\label{se:P22literature}
 While we are not aware of other studies that derived the exact expression \eq{xi22fast} for Eulerian SPT and proposed it for fast numerical evaluation with 1D FFTs, the idea to expand in spherical harmonics and integrate over orientations is a standard method for spherically symmetric problems that can be found throughout the physics literature. 

In the context of large-scale structure, Ref.~\cite{Mccollough1202} obtained related results in Lagrangian perturbation theory (LPT) under the Zel'dovich approximation (ZA), although their derivation does not easily extend to SPT.  In our notation, where the role of upper and lower indices of $\xi^l_n$ is interchanged, Ref.~\cite{Mccollough1202} found for the perturbative 2-2 Zel'dovich correlation function:
 \begin{align}
   \label{eq:xi22_nuala}
   \xi_{(22)}^\mathrm{ZA}(r) = 
\frac{19}{15}[\xi^0_0(r)]^2 
+ \frac{1}{3}\xi^0_{-2}(r)\xi^0_{2}(r)
-\frac{16}{5}\xi^1_{-1}(r)\xi^1_1(r)
+ \frac{34}{21} [\xi^2_0(r)]^2 
+\frac{2}{3}\xi^2_{-2}(r)\xi^2_2(r)
- \frac{4}{5}\xi^3_{-1}(r)\xi^3_1(r)  
+ \frac{4}{35}[\xi^4_0(r)]^2 .
 \end{align}
This expression is similar to our \eqq{xi22fast} for SPT in the sense that it contains the same constituent terms, but the coefficients differ slightly.  This is expected because the Zel'dovich approximation and Eulerian SPT have slightly different  perturbative kernels due to their different approaches to modeling LSS.   A nice feature of Eqs.~\eq{xi22fast} and \eq{xi22_nuala} is that they involve the same $\xi^l_n$'s, so that switching between the Zel'dovich approximation and SPT just involves a change of coefficients.
However, in the case of the Zel'dovich approximation only linear displacements are used, which makes the mapping to the power spectrum particularly simple. 
This also allows for the efficient evaluation of the closed form of the mapping, using again Hankel transforms \cite{zvonimir1410}.

In the context of SPT, Ref.~\cite{sherwinZaldarriaga} provides an expression similar to \eqq{xi22fast} based on a similar configuration space calculation. Their final result however involves gradients of the correlation function that are 
not as computationally efficient as our 1D FFT approach. For $P_{13}$, our result in the next section is exact and differs therefore from other approximate configuration space calculations. Other formally related work that we are aware of includes e.g.~\cite{fergusson0912,fergusson1008,slepian1411,kendrickTrisp1502}.

\section{Fast evaluation of 1-3 power spectrum}
\label{se:Fast13}

\subsection{Main expression}

In configuration space, 1-3 correlations involve three linear fields evaluated at the same location and one linear field evaluated at another location. 
This is different from 2-2 correlations, which involve two linear fields at one and two linear fields at another location. 
In Fourier space, this leads to a different momentum structure for $P_{13}$ compared to $P_{22}$, involving $P_\lin(k)$ and $P_\lin(q)$ instead of $P_\lin(q)$ and $P_\lin(|\vk-\vq|)$. Also, scalar products in the integrand arise from $F_3(\vq,-\vq,\vk)$ instead of $F_2^2(\vq,\vk-\vq)$.   
We will show here that 1-3 correlations still have the form of a 3D convolution that can be evaluated efficiently with 1D FFTs.

The 1-3 correlation between linear and third order density in Eulerian SPT is given by
\begin{eqnarray}
 2\big\la\delta^{(3)}(\vk)\delta_0(-\vk)\big\ra
&=& 2 \int_{\vq,\vp} F_3(\vq,\vp,\vk-\vq-\vp) \big\la  \delta_0(\vq)\delta_0(\vp)\delta_0(\vk-\vq-\vp) \,\delta_0(-\vk) \big\ra\non\\
&=& 
6 (2\pi)^3 P_\lin(k)\int_{\vq}F_3(\vq, -\vq,\vk) P_\lin(q).
\end{eqnarray}
Writing out the symmetric third order $F_3$ kernel, we have\footnote{This agrees with e.g.~\cite{Carrasco1304} if the integration variable is changed from $\vq$ to $-\vq$ for terms that depend on $\vk+\vq$ instead of $\vk-\vq$.  
\eqq{P13From1304} involves the cosine between $\vk$ and $\vq$ because these wavevectors enter as arguments of the $F_3$ kernel. 
}
\begin{align}
\label{eq:P13From1304}
P_{13}(k) = 6P_\lin(k)\int_{\vq}
\frac{P_\lin(q)}{|\vk-\vq|^2}\bigg[&\frac{5k^2}{63} - \frac{11\vk\cdot\vq}{54}
-\frac{k^2(\vk\cdot\vq)^2}{6q^4}
+\frac{19(\vk\cdot\vq)^3}{63q^4}
-\frac{23k^2\vk\cdot\vq}{378q^2}
-\frac{23(\vk\cdot\vq)^2}{378q^2}
+\frac{(\vk\cdot\vq)^3}{9k^2q^2}\bigg].
\end{align}
This integral consists of 3D convolutions of $q^n(\hat\vk\cdot\hat\vq)^lP_\lin(q)$ with the function $f(\vq)=1/q^2$. 
The 3D Fourier transform of the latter is
\begin{align}
  \label{eq:FTof1ovqsq}
  \int_{\vq}e^{-i\vq\cdot\vr}\,\frac{1}{q^2} = \frac{1}{4\pi r},
\end{align}
which corresponds to the solution of the Poisson equation for the potential of a point particle.
Expressing the 3D convolution as a product in configuration space and integrating over orientations then leads to the general identity\footnote{To obtain \eqq{P13generalFastA}, we use \eqq{ScalprodInPls}, introduce $\vp=\vk-\vq$ with a Dirac delta,  decompose this  in plane waves \eq{DiracDeltaInPlaneWaves} and integrate over $\hat\vq$ with \eqq{IntegrateExpLegendreOverAngle}, over $\vp$ with \eqq{FTof1ovqsq}, and over $\hat\vr$ with \eqq{IntegrateExpLegendreOverAngle}.}
\begin{align}
  \label{eq:P13generalFastA}
  \int_\vq \frac{1}{|\vk-\vq|^2}\,q^{n}\,(\hat\vk\cdot\hat\vq)^l\,P_\lin(q)
=
\sum_{l'=0}^l (2l'+1)\,\alpha_{ll'}
\int_0^\infty \d r \,r\,j_{l'}(kr)\,\xi^{l'}_{n}(r).
\end{align}
We therefore obtain for the 1-3 loop correction to the SPT power spectrum 
\begin{empheq}[box=\fbox]{align}
\nonumber
  P_{13}(k) = P_\lin(k)\bigg[
&\frac{67k^2}{189}\int_0^\infty\d r\,r\,j_0(kr)\,\xi^0_0(r)
-\frac{k^4}{3}\int_0^\infty\d
r\,r\,j_0(kr)\,\xi^0_{-2}(r)\non\\
&+\frac{227k^3}{315}\int_0^\infty\d
r\,r\,j_1(kr)\,\xi^1_{-1}(r)
 -\frac{37k}{45}\int_0^\infty\d
r\,r\,j_1(kr)\,\xi^1_{1}(r)\non\\
&-\frac{2k^4}{3}\int_0^\infty\d r\,r\,j_2(kr)\,\xi^2_{-2}(r)
-\frac{46k^2}{189}\int_0^\infty\d r\,r\,j_2(kr)\,\xi^2_{0}(r)\non\\
\label{eq:P13fast_kq_expansion}
&
+\frac{76k^3}{105}\int_0^\infty\d r\,r\,j_3(kr)\,\xi^3_{-1}(r)
+\frac{4k}{15}\int_0^\infty\d r\,r\,j_3(kr)\,\xi^3_{1}(r)
\bigg].
\end{empheq}
This is a fast expression for $P_{13}$ that can be evaluated with sixteen 1D spherical Hankel transforms, eight of which are also needed to compute $P_{22}$.  
We will use this expression for numerical evaluation in the next section.
Similar expressions for $R_n$ integrals relevant for LPT are presented in Appendix~\ref{se:QnRnIntegrals}.

Note that the coefficients in \eqq{P13fast_kq_expansion} can also be read off by decomposing the integrand of $P_{13}$ in Legendre polynomials in $\nu\equiv \hat\vk\cdot\hat\vq$,
\begin{align} 
\nonumber
  P_{13}(k) = P_\lin(k)
\int_{\vq} 
\frac{P_\lin(q)}{|\vk-\vq|^2}
\bigg\{
&\bigg[
\frac{67k^2}{189}-\frac{k^4}{3q^2}\bigg]
\mathsf{P}_0(\nu)
+\bigg[\frac{227 k^3}{315 q}-\frac{37 k q}{45}\bigg]
\mathsf{P}_1(\nu)
\non\\
\label{eq:P13long}
& 
+\bigg[-\frac{2 k^4}{3 q^2}-\frac{46 k^2}{189}\bigg]
\mathsf{P}_2(\nu)
+\bigg[\frac{76 k^3}{105 q}+\frac{4 k q}{15}\bigg]
\mathsf{P}_3(\nu)
\bigg\}.
\end{align}
Also note that a slightly more general form of \eqq{P13generalFastA} is
\begin{align}
  \label{eq:P13generalFastC}
  \int_\vq q^{n_1}|\vk-\vq|^{n_2}\,(\hat\vk\cdot\hat\vq)^l\,P_\lin(q)
=
\sum_{l'=0}^l (2l'+1)\alpha_{ll'}
\,
4\pi\int_0^\infty \d r \,r^2\,j_{l'}(kr)\,\xi^{l'}_{n_1}(r)\,\Xi^0_{n_2}(r),
\end{align}
where
\begin{eqnarray}
  \label{eq:XiDef}
  \Xi^l_n(r) \equiv
 i^l \int_\vq e^{-i\vq\cdot\vr} \,q^n\, \mathsf{P}_l(\hat\vq\cdot\hat\vr)
= \int_0^\infty\frac{\d q}{2\pi^2}\,q^{2+n} j_l(qr).
\end{eqnarray}
This has an analytical solution
\begin{align}
  \label{eq:11}
\Xi^l_n(r) =  \frac{2^n}{\pi^{3/2}}\frac{\Gamma\left(\frac{l+n+3}{2}\right)}{\Gamma\left(\frac{l-n}{2}\right)} \,\frac{1}{r^{3+n}}
\end{align}
if $n<-1$, $n+l>-3$ and $r>0$, which follows from integral 10.22.43 in \cite{NISTDLMF,Olver:2010:NHMF} (also see \cite{Watson66}).

\subsection{Alternative expressions}

The fast expression \eq{P13fast_kq_expansion} for the $P_{13}$ loop
correction in SPT is not unique, because cosines $\nu=\hat\vk\cdot\hat\vq$ in the
integrand of \eqq{P13From1304} can be substituted by noting that
  \begin{align}
    \label{eq:nuIdentity}
\nu = \frac{1}{2}\left(\frac{k}{q}+\frac{q}{k}-\frac{|\vk-\vq|^2}{kq}\right).
  \end{align}
Since the right hand side does not explicitly depend on $\nu$, this
relation can be used to lower the $l$ indices of spherical Bessel
functions $j_l$ in \eqq{P13fast_kq_expansion}, generating expressions
that are equivalent to \eqq{P13fast_kq_expansion} but involve
different constituent terms.

One example to rearrange the $P_{13}$ integrand \eq{P13From1304} is to replace odd
powers of $\nu$ with \eqq{nuIdentity}. In particular, we choose to
replace $\nu^1$ by \eqq{nuIdentity}, and we write $\nu^3=\nu^2\nu$
and replace the single power of $\nu$ there with
\eqq{nuIdentity}. This yields
\begin{align}
  \label{eq:P13alt}
  P_{13}(k) =&\, P_\lin(k)\int_{\vq}\frac{P_\lin(q)}{|\vk-\vq|^2}
\bigg[
-\frac{11}{18}q^2
+\frac{1}{3}q^2\nu^2
-\frac{20}{63}k^2 
+\frac{55}{63}k^2\nu^2
-\frac{23}{126}\frac{k^4}{q^2}
-\frac{2}{21}\frac{k^4}{q^2}\nu^2 
\bigg]\non\\
&
+P_\lin(k) \int_\vq P_\lin(q)
\bigg[
\frac{11}{18}
-\frac{1}{3}\nu^2
+\frac{23}{126}\frac{k^2}{q^2}
-\frac{19}{21}\frac{k^2}{q^2}\nu^2
\bigg].
\end{align}
In the large-scale limit, $k\ll q$, the terms with coefficients $11/18$ and $1/3$ in the first square brackets cancel with the terms with the same coefficients in the second square brackets, implying $P_{13}(k)\sim k^2 P_\lin(k)$ at low $k$.
To obtain a fast expression, we use \eqq{FTof1ovqsq} and 
\begin{align}
  \label{eq:3}
  \int_\vq q^n (\hat\vk\cdot\hat\vq)^l P_\lin(q) = \alpha_{l0}\,\xi^0_n(0).
\end{align}
This leads to 
\begin{align}
  P_{13}(k) = P_\lin(k)\bigg[&
-\frac{1}{2}\int_0^\infty\d r\,r\,j_0(kr)\xi^0_2(r)
+\frac{2}{9}\int_0^\infty\d r\,r\,j_2(kr)\xi^2_2(r)
\non\\
&
-\frac{5k^2}{189}\int_0^\infty\d r\,r\,j_0(kr)\xi^0_0(r)
+\frac{110k^2}{189}\int_0^\infty\d r\,r\,j_2(kr)\xi^2_0(r)
\non\\
&
-\frac{3k^4}{14}\int_0^\infty\d r\,r\,j_0(kr)\xi^0_{-2}(r)
-\frac{4k^4}{63}\int_0^\infty\d r\,r\,j_2(kr)\xi^2_{-2}(r)
\non\\
\label{eq:P13fast_alternative}
&
+\frac{1}{2}\xi^0_0(0)
-\frac{5k^2}{42}\xi^0_{-2}(0)\bigg].
\end{align}
This fast expression only involves even
powers of $k$ and spherical Bessel functions $j_l$ with even $l$. As before,
all $\xi^l_n$'s that enter are also needed for $P_{22}$. 

The last line
of \eqq{P13fast_alternative} contains the mean-squared mass density
and displacement field at zero separation, which only converge if the
matter power spectrum decreases sufficiently fast on small scales.\footnote{Since $P_{13}$ should not change, such divergences would have to cancel with other terms in the first lines of \eqq{P13fast_alternative}. For example, such a cancellation is easy to identify between the two terms with coefficient $1/2$ in \eqq{P13fast_alternative} for $k=0$. We do not discuss this in more detail here because the expression \eq{P13fast_kq_expansion} for $P_{13}$ that we recommend for numerical evaluation does not have such obvious divergences.}
Since cancellations of such potentially divergent terms are numerically problematic, and we did not identify such divergences in \eqq{P13fast_kq_expansion}, we use \eqq{P13fast_kq_expansion} for numerical evaluation rather than \eqq{P13fast_alternative}.

Note that the coefficients in \eqq{P13fast_alternative} can also be read off by rewriting \eqq{P13alt} in terms of Legendre polynomials,
\begin{align}
  P_{13}(k) =\,& P_\lin(k)\int_{\vq}\frac{P_\lin(q)}{|\vk-\vq|^2}
\bigg[
-\frac{q^2}{2}\mathsf{P}_0(\nu) 
+\frac{2q^2}{9}\mathsf{P}_2(\nu)
-\frac{5k^2}{189}\mathsf{P}_0(\nu)
+\frac{110k^2}{189}\mathsf{P}_2(\nu)
-\frac{3k^4}{14q^2}\mathsf{P}_0(\nu)
-\frac{4k^4}{63q^2}\mathsf{P}_2(\nu)
\bigg]\non\\
&
+P_\lin(k) \int_{\vq}P_\lin(q)\bigg[
\frac{1}{2}\mathsf{P}_0(\nu)
-\frac{2}{9}\mathsf{P}_2(\nu)
-\frac{5k^2}{42q^2}\mathsf{P}_0(\nu)
+\frac{38k^2}{63q^2}\mathsf{P}_2(\nu)
\bigg].
\end{align}

\section{Numerical evaluation}
\label{se:numerics}

\begin{figure}[thbp]
\centerline{
\includegraphics[width=0.7\textwidth]{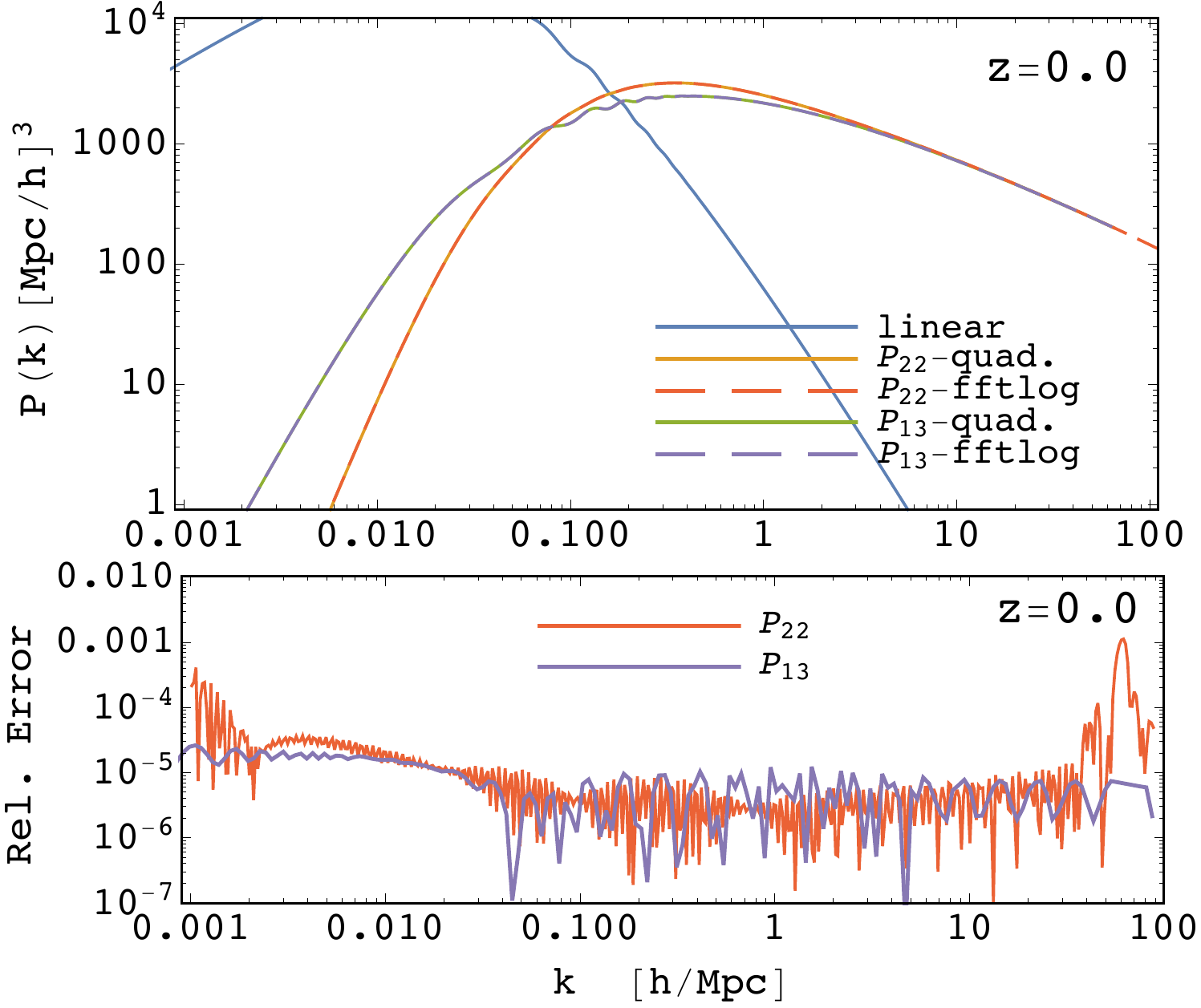}}
\caption{Loop corrections $P_{22}$ and $P_{13}$ to the SPT matter power spectrum computed with conventional quadrature integration of the convolution integrals \eq{P22_def} and \eq{P13From1304}, and with the new 1D FFT method of Eqs.~\eq{P22fromxi22}, \eq{xi22fast} and \eq{P13fast_kq_expansion}.  
The bottom panel shows the relative error between the two methods.
Its size is dominated by the accuracy of the quadrature routine which was set to $10^{-5}$.
}
\label{fig:P22andP13}
\end{figure}

Fig.~\ref{fig:P22andP13} shows $P_{22}$ and $P_{13}$ computed with Eqs.~\eq{P22fromxi22}, \eq{xi22fast} and \eq{P13fast_kq_expansion} using only fast 1D Hankel transforms evaluated with  FFTLog \cite{hamiltonfftlog}.  This is compared against the traditional method of performing the convolution integrals \eq{P22_def} and \eq{P13From1304}, which involves a 2D quadrature integration for every $k$ in the case of $P_{22}$, and a 1D quadrature integration for every $k$ in the case of $P_{13}$.
The relative agreement of the methods is better than $0.1\%$ over five orders in magnitude in scales, $0.001\,h/\mathrm{Mpc}\le k \le 100\,h/\mathrm{Mpc}$.
On scales where loop corrections are most relevant, the fractional difference between the methods is around $10^{-5}$.
This fractional difference is dominated by the fact that we set the accuracy of the quadrature integration routine to $10^{-5}$. 
The fractional deviation of the FFT result from the true integration result should thus be less than $10^{-5}$.
In principle it should be exact up to machine level precision.

This demonstrates that the new 1D FFT/Hankel transform method to compute 1-loop power spectra  is numerically stable and extremely accurate.

\section{Conclusions}
\label{se:conclusions}

Nonlinear corrections to models for the large-scale structure of the universe are becoming an increasingly important ingredient for future large-scale structure surveys. 
In this paper, we establish a new methodology to numerically evaluate such nonlinear corrections more efficiently than commonly used methods.
In particular, we derived new expressions for the 2-2 and 1-3 loop contributions to the matter power spectrum in Eulerian Standard Perturbation Theory, given by Eqs.~\eq{P22fromxi22}, \eq{xi22fast} and \eq{P13fast_kq_expansion}, and for loop integrals relevant for Lagrangian Perturbation Theory, listed in Appendix~\ref{se:QnRnIntegrals}.
These new expressions can be evaluated at all output wavenumbers $k$ with just a few one-dimensional FFTs in an exact manner.
In our implementation we found excellent agreement with commonly used quadrature integration methods.
At the same time, the method is a few orders of magnitude faster because it exploits FFTs.

We expect this numerical speedup to be useful for various practical use cases where large-scale structure perturbation theory is currently applied.
For example, it can speed up Monte Carlo chains to explore the cosmological parameter space when interpreting large-scale structure observations.
The FFT method can in principle also be applied to higher order corrections to the power spectrum where brute-force quadrature integration becomes prohibitively expensive.

The fast expressions for loop corrections to the matter power spectrum can be derived by working in configuration space instead of the more commonly used Fourier space.
The fact that the perturbative expansion of the density involves products of fields in configuration space simplifies the calculation compared to Fourier space, where configuration space products become convolutions.
An alternative derivation of our results follows by evaluating three-dimensional convolution integrals as Fourier transforms of products in configuration space (and exploiting spherical symmetry of the problem).
We derive general identities for three-dimensional convolution integrals given by Eqs.~\eq{P22FormFinal}, \eq{P22FormFinal2}, \eq{P13generalFastA} and \eq{P13generalFastC}. 
These may be useful in other contexts where formally similar integrals appear.

We expect that the approach proposed in this work can be extended in various ways, e.g.~to halos in redshift space or higher loop corrections and higher order N-point functions.  
Given the plethora of upcoming large-scale structure experiments and cosmological inferences possible with these datasets, such extensions should be very worthwhile.

Code developed for this paper to evaluate 1-loop power spectra with 1D FFTs can be obtained from the authors upon request by emailing \href{mailto:zvlah@stanford.edu}{zvlah@stanford.edu}.

\section*{Acknowledgments}
We thank Uro\v{s} Seljak and Martin White for useful discussions.
Z.V. is supported in part by the U.S. Department of Energy contract to SLAC no. DE-AC02-76SF00515.

\appendix

\section{Configuration space evolution equations}
\label{se:configspaceevolution}

Using the usual well-motivated approximation that the velocity field is 
irrotational, and defining the 
velocity divergence $\theta \equiv \vnabla \cdot \vv$, and velocity 
potential $\nabla ^2 \psi \equiv \theta$, we can, with simple vector calculus
manipulations, write the fluid evolution 
equations in the following form, assuming an Einstein-de Sitter Universe
for simplicity:
\begin{equation}
\frac{\partial \delta}{\partial \tau}+\theta=
\frac{1}{2}\left[\psi\nabla^2\delta-\delta \theta - 
\nabla^2\left(\delta \psi\right)\right]
\end{equation}
and
\begin{equation}
\frac{\partial \theta}{\partial \tau}+
\mathcal{H} \theta+
\frac{3}{2}\mathcal{H}^2 \delta=
\frac{1}{2}\left[\nabla^2(\psi \theta)-\frac{1}{2}\nabla^4\psi^2\right]
\end{equation}
where $\tau$ is the conformal time defined by $d \tau=dt/a(t)$, and
$\mathcal{H}\equiv d\ln a /d\tau = H a$. This form makes explicit the fact
that this evolution is local in $\delta(\vx)$, $\theta(\vx)$, and 
$\psi(\vx)$ (counting Laplacians as local, as they involve only differences in
the field at infinitesimal separation). The evolution is of course not 
actually local in
the evolution variables, density and velocity, as evolving them requires 
computing the integral $\psi= \nabla^{-2}\theta$, but we are used to doing this
kind of integral efficiently as a multiplication by $k^{-2}$ in Fourier space. 
While we do not use this approach explicitly in this paper, this form of the
equations motivates writing the usual Eulerian SPT recursion relations in 
configuration space in terms of these fields, which can be done to produce 
expressions exactly equivalent to the usual integrals over Fourier space 
kernels like $F_2$, $F_3$, etc.. From there one can derive expressions for,
e.g., $\xi_{22}(r)$, entirely as products of correlations of the above 
fields, and Laplacians of these correlation functions. This motivates the 
idea that we can evaluate PT terms using a combination of products in 
configuration space and derivatives in Fourier space, with FFTLog 
transformations between them, similar to the approach used in spectral method 
numerical simulations. 
The various different Hankel transforms that we use in
the main text
turn out to be a concise way to evaluate the combinations of derivatives that
would otherwise appear in the higher order correlation function expressions. 

\section{Fast expressions for $Q_n$ and $R_n$ integrals in Lagrangian Perturbation Theory
(LPT)}
\label{se:QnRnIntegrals}

Integrals similar to $P_{22}$ and $P_{13}$ often appear in the literature. Here we provide fast expressions for  integrals that arise when computing polyspectra of the displacement field that play an important role in Lagrangian Perturbation Theory (LPT). Since the relevant displacement kernels $L^{(n)}$ involve the same terms as the $F^{(n)}$ kernels, the integrals contain the same type of terms as SPT 1-loop power spectra but with modified coefficients. 

\subsection{Fast $Q_n$ integrals}

We use the same definitions as in \cite{Matsubara0807} for all $Q_n$ integrals, e.g.
\begin{align}
  \label{eq:15}
  Q_1 (k) &= \frac{k^3}{4\pi^2} \int_0^\infty \d r\,P_\lin(kr)\int_{-1}^1\d x\,P_\lin(k(1+r^2-2rx)^{1/2})\,\frac{r^2(1-x^2)^2}{(1+r^2-2rx)^2}\\
\label{eq:Q1_with_q}
&= k^4\int_{\vq}\frac{[1-(\hat\vk\cdot\hat\vq)^2]^2}{|\vk-\vq|^4} P_\lin(q)P_\lin(|\vk-\vq|),
\end{align}
where $q=kr$ and $x=\hat\vk\cdot\hat\vq$, so that $rx=\vk\cdot\vq/k^2$ and
\begin{align}
  \label{eq:14}
  \frac{k^3}{4\pi^2}\int_0^\infty \d r\,\int_{-1}^1\d x\,\cdots \;=\; \int\frac{\d^3\vq}{(2\pi)^3}\frac{k^2}{q^2}\cdots.
\end{align}
Rather than parameterizing the orientation of the integration variable $\vq$  in terms of $x=\hat\vq\cdot\hat\vk$, it can be parameterized in terms of $\mu=\hat\vq\cdot(\widehat{\vk-\vq})$.
Then, all $Q_n$ integrals from \cite{Matsubara0807} can be expressed in terms of convolution integrals of the general form given by \eqq{P22FormFinal2}.\footnote{To convert the integrands from Ref.~\cite{Matsubara0807}, note that $1+r^2-2rx=|\vk-\vq|^2/k^2$ and
  \begin{align}
    \label{eq:26}
1-x^2=\frac{|\vk-\vq|^2}{k^2}(1-\mu^2),
\qquad
    x^2 = \frac{q^2}{k^2}\left(1+\frac{|\vk-\vq|}{q}\mu\right)^2,
\qquad
1-rx = \frac{|\vk-\vq|^2}{k^2}\left(1+\frac{q}{|\vk-\vq|}\,\mu\right),
\qquad
rx = \frac{q^2}{k^2}\left(1+\frac{|\vk-\vq|}{q}\mu\right).
  \end{align}
Powers of $k$ can be mitigated by noting that $k^2=(\vk-\vq+\vq)^2$ implies
\begin{align}
  \label{eq:kn_in_terms_of_pqmu}
  \frac{k^2}{q|\vk-\vq|} = \frac{|\vk-\vq|}{q}+\frac{q}{|\vk-\vq|}+2\mu.
\end{align}
}

To apply this to $Q_1$ we express it in terms of $\mu$,
\begin{eqnarray}
  \label{eq:23}
  Q_1(k)
 &=& 
 \int_\vq\frac{[k^2q^2-(\vk\cdot\vq)^2]^2}{q^4|\vk-\vq|^4}P_\lin(q)P_\lin(|\vk-\vq|)\\
&=& \int_\vq 
(1-2\mu^2+\mu^4)
P_\lin(q)P_\lin(|\vk-\vq|).
\end{eqnarray}
Using \eqq{P22FormFinal2} this integral becomes 
\begin{align}
  \label{eq:19}
  Q_1(k) = 4\pi\int_0^\infty\d r\,r^2 j_0(kr)\sum_{l'=0}^4 (-1)^{l'} (2l'+1)\left\{
\alpha_{0,l'}[\xi^{l'}_0(r)]^2
-2\alpha_{2,l'}[\xi^{l'}_0(r)]^2
+\alpha_{4,l'}[\xi^{l'}_0(r)]^2
\right\},
\end{align}
where we extended all sums up to $l'=4$, which does not change the result because $\alpha_{ll'}=0$ for $l'>l$.
Plugging in $\alpha$'s,
\begin{align}
  \label{eq:Q1fast}
Q_1(k) 
= 4\pi\int_0^\infty\d r\,r^2j_0(kr) \left\{
\frac{8}{15}[\xi^0_0(r)]^2 -\frac{16}{21}[\xi^2_0(r)]^2 + \frac{8}{35}[\xi^4_0(r)]^2 
\right\}.
\end{align}
This is a  $l=0$ 1D Hankel transform of a sum of products of $\xi^l_n$'s which is fast to evaluate. The same terms also arise when computing $P_{22}$ with \eqq{xi22fast}.

 Similar fast expressions follow for the other $Q_n$ integrals, e.g.
\begin{eqnarray}
  \label{eq:25}
  Q_2(k) &=& \int_\vq\left\{
\left(1-\mu^2\right) \left[1+\left(\frac{q}{|\vk-\vq|}+\frac{|\vk-\vq|}{q}\right)\mu+\mu^2\right] \right\} P_\lin(q)P_\lin(|\vk-\vq|)\\
&=& 4\pi\int_0^\infty\d r\,r^2j_0(kr)\left\{
\frac{4}{5}[\xi^0_0(r)]^2 - \frac{4}{7}[\xi^2_0(r)]^2 - \frac{8}{35}[\xi^4_0(r)]^2
 - \frac{4}{5}\xi^1_1(r)\xi^1_{-1}(r) + \frac{4}{5}\xi^3_1(r)\xi^3_{-1}(r)
\right\},
\\
Q_3(k) &=& \int_\vq \left[1+\left(\frac{q}{|\vk-\vq|}+\frac{|\vk-\vq|}{q}\right)\mu+\mu^2\right]^2P_\lin(q)P_\lin(|\vk-\vq|)\\
&=& 4\pi\int_0^\infty\d r\,r^2j_0(kr) \bigg\{
\frac{38}{15}[\xi^0_0(r)]^2 + \frac{2}{3}\xi^0_{-2}(r)\xi^0_2(r) 
- \frac{32}{5}\xi^1_{-1}(r)\xi^1_1(r)  + \frac{68}{21}[\xi^2_0(r)]^2 \non\\
&&\qquad\qquad\qquad\qquad\quad + \frac{4}{3}\xi^2_{-2}(r)\xi^2_{2}(r)
-\frac{8}{5}\xi^3_{-1}(r)\xi^3_1(r)
+\frac{8}{35}[\xi^4_0(r)]^2
\bigg\}.
\end{eqnarray}
As a consistency check, we confirmed that $9Q_1/98  + 3Q_2/7  +  Q_3/2=P_{22}$ is satisfied \cite{Matsubara:2007wj, zvonimir1410}, 
i.e.~the coefficients in $Q_1$, $Q_2$ and $Q_3$ add up to those of $P_{22}$ in \eqq{xi22fast}.

\subsection{Fast $R_n$ integrals}

Similarly, the $R_n$ integrals of \cite{Matsubara:2007wj, Matsubara0807} are related to $P_{13}$ integrals and fast expressions can be obtained similarly to $P_{13}$. For example, 
\begin{eqnarray}
  \label{eq:9}
  R_1(k) &=& \frac{k^3}{4\pi^2} P_\lin(k) \int_0^\infty \d r\, P_\lin(kr) \int_{-1}^1\d x\,
\frac{r^2(1-x^2)^2}{1+r^2-2rx} \\
&=& P_\lin(k) \int_\vq \frac{k^2}{|\vk-\vq|^2}(1-\nu^2)^2 P_\lin(q) \\
\label{eq:R1fast}
&=& k^2P_\lin(k) \left[
\frac{8}{15}\int_0^\infty\d r\, r\, j_0(kr)\xi^0_0(r)
-\frac{16}{21}\int_0^\infty\d r\, r\, j_2(kr)\xi^2_0(r)
+\frac{8}{35}\int_0^\infty\d r\, r\, j_4(kr)\xi^4_0(r)
\right],
\end{eqnarray}
where $\nu=x=\hat\vq\cdot\hat\vk$, and we used \eqq{P13generalFastA} for the last step.
Due to related angular structures, \eqq{R1fast} for $R_1$ and \eqq{Q1fast} for $Q_1$ involve the same coefficients.

For the $R_2$ integral of \cite{Matsubara:2007wj, Matsubara0807} we obtain
\begin{eqnarray}
  \label{eq:4}
  R_2(k) &=& P_\lin(k)\int_\vq P_\lin(q) \frac{1}{|\vk-\vq|^2}
\left[
\frac{k^3}{q}(\nu-\nu^3) + k^2 (\nu^4-\nu^2)
\right]\\
&=& P_\lin(k)\bigg[
-\frac{2k^2}{15} \int \d r\,r \,j_0(kr)\xi^0_0(r)
-\frac{2k^2}{21} \int \d r\,r \,j_2(kr)\xi^2_0(r)
+\frac{8k^2}{35} \int \d r\,r \,j_4(kr)\xi^4_0(r)\non\\
&&\qquad \qquad
+\frac{2k^3}{5} \int \d r\,r \,j_1(kr)\xi^1_{-1}(r)
-\frac{2k^3}{5} \int \d r\,r \,j_3(kr)\xi^3_{-1}(r)\bigg].
\end{eqnarray}

We again check if the relation $10R_1/21  + 6R_2/7  - 2 k^2 \sigma^2 P_\lin=P_{13}$ 
is satisfied \cite{Matsubara:2007wj, zvonimir1410}, where $\sigma^2 = \frac{1}{3} \int_\vq P_\lin(q)/q^2$.
Contrary to the $P_{22}$ case, where simply adding up the $Q_1$, $Q_2$ and $Q_3$ coefficients is sufficient to 
check the consistency, in the case of $P_{13}$ recurrence relations of the Bessel functions have to be used.
Alternatively, these consistency checks for both $P_{22}$ and $P_{13}$ can also be performed numerically.

\section{Useful mathematical identities}
This section  lists some useful mathematical identities for calculations in the main text.

The addition theorem for spherical harmonics is
\begin{equation}
  \label{eq:YlmAdditionThm}
  \mathsf{P}_l(\hat{\vq}_1\cdot \hat{\vq}_2) = \frac{4\pi}{2l+1}
\sum_{m=-l}^l Y_{lm}(\hat{\vq}_1)Y_{lm}^*(\hat{\vq}_2).
\end{equation}
The angular part of a 3D Fourier transform of a spherical harmonic yields (using Eqs.~\eq{ExpInLegendres} and \eq{YlmAdditionThm})
\begin{align}
  \label{eq:FTofYlm}
  \int\d\Omega_{\hat\vq}\,e^{\pm ia\vq\cdot\vr}\,Y_{lm}(\hat\vq) = 4\pi (\pm i\,\sgn\, a)^l j_l(|a|qr)Y_{lm}(\hat \vr).
\end{align}
The Dirac delta can be expanded in plane waves,
\begin{align}
  \label{eq:DiracDeltaInPlaneWaves}
  (2\pi)^3\delta_D(\vq) = \int\d^3 \vr\,e^{i\vq\cdot\vr}.
\end{align}
The plane wave expansion for an arbitrary parameter $a\in\mathbb{R}$ is
\begin{align}
  \label{eq:ExpInLegendres}
  e^{\pm i a\vk\cdot\vr} = \sum_{l=0}^\infty (2l+1)(\pm i\, \sgn(a))^l\, j_l(|a|kr)\,\mathsf{P}_l(\hat\vk\cdot\hat\vr).
\end{align}
This includes the case $a=0$ for which $j_l(0)=\delta_{l0}$.
From \eqq{YlmAdditionThm},
\begin{align}
  \label{eq:ExpInYlms}
  e^{\pm i a\vk\cdot\vr} = 4\pi \sum_{l=0}^\infty\sum_{m=-l}^l (\pm i\, \sgn(a))^l\, j_l(|a|kr)\,Y_{lm}(\hat\vk)Y^*_{lm}(\hat\vr).
\end{align}

The addition theorem for spherical harmonics \eq{YlmAdditionThm} implies the following orthogonality relation for the integral over two Legendre polynomials
\begin{align}
  \label{eq:Int2Legendres}
 \int \d\Omega_{\hat\vq_2}\, \mathsf{P}_l(\hat{\vq}_1\cdot\hat{\vq}_2)\mathsf{P}_{l'}(\hat\vq_2\cdot\hat\vq_3)
=
\delta_{ll'}\frac{4\pi}{2l+1}\mathsf{P}_l({\hat\vq_1\cdot\hat\vq_3}).
\end{align}
Eqs.~\eq{ExpInLegendres} and \eq{Int2Legendres} can be used to perform the angular part of the 3D Fourier transform of a Legendre polynomial,
\begin{align}
  \label{eq:IntegrateExpLegendreOverAngle}
  \int\d\Omega_{\hat\vq}\, e^{\pm ia\vq\cdot\vr}\,\mathsf{P}_l(\hat\vq\cdot\hat\vk)
= 4\pi (\pm i\,\sgn(a))^l \,j_l(|a|qr)\, \mathsf{P}_l(\hat\vk\cdot\hat\vr),
\end{align}
where $ a\in\mathbb{R}$.

\bibliography{bib_fft_loops}

\begin{thebibliography}{35}%
\makeatletter
\providecommand \@ifxundefined [1]{%
 \@ifx{#1\undefined}
}%
\providecommand \@ifnum [1]{%
 \ifnum #1\expandafter \@firstoftwo
 \else \expandafter \@secondoftwo
 \fi
}%
\providecommand \@ifx [1]{%
 \ifx #1\expandafter \@firstoftwo
 \else \expandafter \@secondoftwo
 \fi
}%
\providecommand \natexlab [1]{#1}%
\providecommand \enquote  [1]{``#1''}%
\providecommand \bibnamefont  [1]{#1}%
\providecommand \bibfnamefont [1]{#1}%
\providecommand \citenamefont [1]{#1}%
\providecommand \href@noop [0]{\@secondoftwo}%
\providecommand \href [0]{\begingroup \@sanitize@url \@href}%
\providecommand \@href[1]{\@@startlink{#1}\@@href}%
\providecommand \@@href[1]{\endgroup#1\@@endlink}%
\providecommand \@sanitize@url [0]{\catcode `\\12\catcode `\$12\catcode
  `\&12\catcode `\#12\catcode `\^12\catcode `\_12\catcode `\%12\relax}%
\providecommand \@@startlink[1]{}%
\providecommand \@@endlink[0]{}%
\providecommand \url  [0]{\begingroup\@sanitize@url \@url }%
\providecommand \@url [1]{\endgroup\@href {#1}{\urlprefix }}%
\providecommand \urlprefix  [0]{URL }%
\providecommand \Eprint [0]{\href }%
\providecommand \doibase [0]{http://dx.doi.org/}%
\providecommand \selectlanguage [0]{\@gobble}%
\providecommand \bibinfo  [0]{\@secondoftwo}%
\providecommand \bibfield  [0]{\@secondoftwo}%
\providecommand \translation [1]{[#1]}%
\providecommand \BibitemOpen [0]{}%
\providecommand \bibitemStop [0]{}%
\providecommand \bibitemNoStop [0]{.\EOS\space}%
\providecommand \EOS [0]{\spacefactor3000\relax}%
\providecommand \BibitemShut  [1]{\csname bibitem#1\endcsname}%
\let\auto@bib@innerbib\@empty
%</preamble>
\bibitem [{\citenamefont {{The Dark Energy Survey
  Collaboration}}(2005)}]{DESwhitepaper}%
  \BibitemOpen
  \bibfield  {author} {\bibinfo {author} {\bibnamefont {{The Dark Energy Survey
  Collaboration}}},\ }\href@noop {} {\bibfield  {journal} {\bibinfo  {journal}
  {ArXiv Astrophysics e-prints}\ } (\bibinfo {year} {2005})},\ \Eprint
  {http://arxiv.org/abs/astro-ph/0510346} {astro-ph/0510346} \BibitemShut
  {NoStop}%
\bibitem [{\citenamefont {{Dawson}}\ \emph {et~al.}(2016)\citenamefont
  {{Dawson}}, \citenamefont {{Kneib}}, \citenamefont {{Percival}},
  \citenamefont {{Alam}}, \citenamefont {{Albareti}}, \citenamefont
  {{Anderson}}, \citenamefont {{Armengaud}}, \citenamefont {{Aubourg}},
  \citenamefont {{Bailey}}, \citenamefont {{Bautista}} \emph
  {et~al.}}]{eBOSSDawson}%
  \BibitemOpen
  \bibfield  {author} {\bibinfo {author} {\bibfnamefont {K.~S.}\ \bibnamefont
  {{Dawson}}}, \bibinfo {author} {\bibfnamefont {J.-P.}\ \bibnamefont
  {{Kneib}}}, \bibinfo {author} {\bibfnamefont {W.~J.}\ \bibnamefont
  {{Percival}}}, \bibinfo {author} {\bibfnamefont {S.}~\bibnamefont {{Alam}}},
  \bibinfo {author} {\bibfnamefont {F.~D.}\ \bibnamefont {{Albareti}}},
  \bibinfo {author} {\bibfnamefont {S.~F.}\ \bibnamefont {{Anderson}}},
  \bibinfo {author} {\bibfnamefont {E.}~\bibnamefont {{Armengaud}}}, \bibinfo
  {author} {\bibfnamefont {{\'E}.}~\bibnamefont {{Aubourg}}}, \bibinfo {author}
  {\bibfnamefont {S.}~\bibnamefont {{Bailey}}}, \bibinfo {author}
  {\bibfnamefont {J.~E.}\ \bibnamefont {{Bautista}}},  \emph {et~al.},\ }\href
  {\doibase 10.3847/0004-6256/151/2/44} {\bibfield  {journal} {\bibinfo
  {journal} {The Astronomical Journal}\ }\textbf {\bibinfo {volume} {151}},\
  \bibinfo {eid} {44} (\bibinfo {year} {2016})},\ \Eprint
  {http://arxiv.org/abs/1508.04473} {arXiv:1508.04473} \BibitemShut {NoStop}%
\bibitem [{\citenamefont {{Levi}}\ \emph {et~al.}(2013)\citenamefont {{Levi}},
  \citenamefont {{Bebek}}, \citenamefont {{Beers}}, \citenamefont {{Blum}},
  \citenamefont {{Cahn}}, \citenamefont {{Eisenstein}}, \citenamefont
  {{Flaugher}}, \citenamefont {{Honscheid}}, \citenamefont {{Kron}},
  \citenamefont {{Lahav}}, \citenamefont {{McDonald}}, \citenamefont {{Roe}},
  \citenamefont {{Schlegel}},\ and\ \citenamefont {{representing the DESI
  collaboration}}}]{DESIwhitepaper}%
  \BibitemOpen
  \bibfield  {author} {\bibinfo {author} {\bibfnamefont {M.}~\bibnamefont
  {{Levi}}}, \bibinfo {author} {\bibfnamefont {C.}~\bibnamefont {{Bebek}}},
  \bibinfo {author} {\bibfnamefont {T.}~\bibnamefont {{Beers}}}, \bibinfo
  {author} {\bibfnamefont {R.}~\bibnamefont {{Blum}}}, \bibinfo {author}
  {\bibfnamefont {R.}~\bibnamefont {{Cahn}}}, \bibinfo {author} {\bibfnamefont
  {D.}~\bibnamefont {{Eisenstein}}}, \bibinfo {author} {\bibfnamefont
  {B.}~\bibnamefont {{Flaugher}}}, \bibinfo {author} {\bibfnamefont
  {K.}~\bibnamefont {{Honscheid}}}, \bibinfo {author} {\bibfnamefont
  {R.}~\bibnamefont {{Kron}}}, \bibinfo {author} {\bibfnamefont
  {O.}~\bibnamefont {{Lahav}}}, \bibinfo {author} {\bibfnamefont
  {P.}~\bibnamefont {{McDonald}}}, \bibinfo {author} {\bibfnamefont
  {N.}~\bibnamefont {{Roe}}}, \bibinfo {author} {\bibfnamefont
  {D.}~\bibnamefont {{Schlegel}}}, \ and\ \bibinfo {author} {\bibnamefont
  {{representing the DESI collaboration}}},\ }\href@noop {} {\bibfield
  {journal} {\bibinfo  {journal} {ArXiv e-prints}\ } (\bibinfo {year}
  {2013})},\ \Eprint {http://arxiv.org/abs/1308.0847} {arXiv:1308.0847
  [astro-ph.CO]} \BibitemShut {NoStop}%
\bibitem [{\citenamefont {{Laureijs}}\ \emph {et~al.}(2011)\citenamefont
  {{Laureijs}}, \citenamefont {{Amiaux}}, \citenamefont {{Arduini}},
  \citenamefont {{Augu{\`e}res}}, \citenamefont {{Brinchmann}}, \citenamefont
  {{Cole}}, \citenamefont {{Cropper}}, \citenamefont {{Dabin}}, \citenamefont
  {{Duvet}}, \citenamefont {{Ealet}} \emph {et~al.}}]{EuclidWhitePaper}%
  \BibitemOpen
  \bibfield  {author} {\bibinfo {author} {\bibfnamefont {R.}~\bibnamefont
  {{Laureijs}}}, \bibinfo {author} {\bibfnamefont {J.}~\bibnamefont
  {{Amiaux}}}, \bibinfo {author} {\bibfnamefont {S.}~\bibnamefont {{Arduini}}},
  \bibinfo {author} {\bibfnamefont {J.~.}\ \bibnamefont {{Augu{\`e}res}}},
  \bibinfo {author} {\bibfnamefont {J.}~\bibnamefont {{Brinchmann}}}, \bibinfo
  {author} {\bibfnamefont {R.}~\bibnamefont {{Cole}}}, \bibinfo {author}
  {\bibfnamefont {M.}~\bibnamefont {{Cropper}}}, \bibinfo {author}
  {\bibfnamefont {C.}~\bibnamefont {{Dabin}}}, \bibinfo {author} {\bibfnamefont
  {L.}~\bibnamefont {{Duvet}}}, \bibinfo {author} {\bibfnamefont
  {A.}~\bibnamefont {{Ealet}}},  \emph {et~al.},\ }\href@noop {} {\bibfield
  {journal} {\bibinfo  {journal} {ArXiv e-prints}\ } (\bibinfo {year}
  {2011})},\ \Eprint {http://arxiv.org/abs/1110.3193} {arXiv:1110.3193
  [astro-ph.CO]} \BibitemShut {NoStop}%
\bibitem [{\citenamefont {{Spergel}}\ \emph {et~al.}(2015)\citenamefont
  {{Spergel}}, \citenamefont {{Gehrels}}, \citenamefont {{Baltay}},
  \citenamefont {{Bennett}}, \citenamefont {{Breckinridge}}, \citenamefont
  {{Donahue}}, \citenamefont {{Dressler}}, \citenamefont {{Gaudi}},
  \citenamefont {{Greene}}, \citenamefont {{Guyon}}, \citenamefont {{Hirata}}
  \emph {et~al.}}]{WFIRST1503}%
  \BibitemOpen
  \bibfield  {author} {\bibinfo {author} {\bibfnamefont {D.}~\bibnamefont
  {{Spergel}}}, \bibinfo {author} {\bibfnamefont {N.}~\bibnamefont
  {{Gehrels}}}, \bibinfo {author} {\bibfnamefont {C.}~\bibnamefont {{Baltay}}},
  \bibinfo {author} {\bibfnamefont {D.}~\bibnamefont {{Bennett}}}, \bibinfo
  {author} {\bibfnamefont {J.}~\bibnamefont {{Breckinridge}}}, \bibinfo
  {author} {\bibfnamefont {M.}~\bibnamefont {{Donahue}}}, \bibinfo {author}
  {\bibfnamefont {A.}~\bibnamefont {{Dressler}}}, \bibinfo {author}
  {\bibfnamefont {B.~S.}\ \bibnamefont {{Gaudi}}}, \bibinfo {author}
  {\bibfnamefont {T.}~\bibnamefont {{Greene}}}, \bibinfo {author}
  {\bibfnamefont {O.}~\bibnamefont {{Guyon}}}, \bibinfo {author} {\bibfnamefont
  {C.}~\bibnamefont {{Hirata}}},  \emph {et~al.},\ }\href@noop {} {\bibfield
  {journal} {\bibinfo  {journal} {ArXiv e-prints}\ } (\bibinfo {year}
  {2015})},\ \Eprint {http://arxiv.org/abs/1503.03757} {arXiv:1503.03757
  [astro-ph.IM]} \BibitemShut {NoStop}%
\bibitem [{\citenamefont {{LSST Dark Energy Science
  Collaboration}}(2012)}]{LSSTDESC}%
  \BibitemOpen
  \bibfield  {author} {\bibinfo {author} {\bibnamefont {{LSST Dark Energy
  Science Collaboration}}},\ }\href@noop {} {\bibfield  {journal} {\bibinfo
  {journal} {ArXiv e-prints}\ } (\bibinfo {year} {2012})},\ \Eprint
  {http://arxiv.org/abs/1211.0310} {arXiv:1211.0310 [astro-ph.CO]} \BibitemShut
  {NoStop}%
\bibitem [{\citenamefont {{Dor{\'e}}}\ \emph {et~al.}(2014)\citenamefont
  {{Dor{\'e}}}, \citenamefont {{Bock}}, \citenamefont {{Ashby}}, \citenamefont
  {{Capak}}, \citenamefont {{Cooray}}, \citenamefont {{de Putter}},
  \citenamefont {{Eifler}}, \citenamefont {{Flagey}}, \citenamefont {{Gong}},
  \citenamefont {{Habib}} \emph {et~al.}}]{spherex1412}%
  \BibitemOpen
  \bibfield  {author} {\bibinfo {author} {\bibfnamefont {O.}~\bibnamefont
  {{Dor{\'e}}}}, \bibinfo {author} {\bibfnamefont {J.}~\bibnamefont {{Bock}}},
  \bibinfo {author} {\bibfnamefont {M.}~\bibnamefont {{Ashby}}}, \bibinfo
  {author} {\bibfnamefont {P.}~\bibnamefont {{Capak}}}, \bibinfo {author}
  {\bibfnamefont {A.}~\bibnamefont {{Cooray}}}, \bibinfo {author}
  {\bibfnamefont {R.}~\bibnamefont {{de Putter}}}, \bibinfo {author}
  {\bibfnamefont {T.}~\bibnamefont {{Eifler}}}, \bibinfo {author}
  {\bibfnamefont {N.}~\bibnamefont {{Flagey}}}, \bibinfo {author}
  {\bibfnamefont {Y.}~\bibnamefont {{Gong}}}, \bibinfo {author} {\bibfnamefont
  {S.}~\bibnamefont {{Habib}}},  \emph {et~al.},\ }\href@noop {} {\bibfield
  {journal} {\bibinfo  {journal} {ArXiv e-prints}\ } (\bibinfo {year}
  {2014})},\ \Eprint {http://arxiv.org/abs/1412.4872} {arXiv:1412.4872}
  \BibitemShut {NoStop}%
\bibitem [{\citenamefont {Goroff}\ \emph {et~al.}(1986)\citenamefont {Goroff},
  \citenamefont {Grinstein}, \citenamefont {Rey},\ and\ \citenamefont
  {Wise}}]{Goroff:1986ep}%
  \BibitemOpen
  \bibfield  {author} {\bibinfo {author} {\bibfnamefont {M.~H.}\ \bibnamefont
  {Goroff}}, \bibinfo {author} {\bibfnamefont {B.}~\bibnamefont {Grinstein}},
  \bibinfo {author} {\bibfnamefont {S.~J.}\ \bibnamefont {Rey}}, \ and\
  \bibinfo {author} {\bibfnamefont {M.~B.}\ \bibnamefont {Wise}},\ }\href
  {\doibase 10.1086/164749} {\bibfield  {journal} {\bibinfo  {journal}
  {Astrophys. J.}\ }\textbf {\bibinfo {volume} {311}},\ \bibinfo {pages} {6}
  (\bibinfo {year} {1986})}\BibitemShut {NoStop}%
%%CITATION = ASJOA,311,6;%%
\bibitem [{\citenamefont {Jain}\ and\ \citenamefont
  {Bertschinger}(1994)}]{Jain:1993jh}%
  \BibitemOpen
  \bibfield  {author} {\bibinfo {author} {\bibfnamefont {B.}~\bibnamefont
  {Jain}}\ and\ \bibinfo {author} {\bibfnamefont {E.}~\bibnamefont
  {Bertschinger}},\ }\href {\doibase 10.1086/174502} {\bibfield  {journal}
  {\bibinfo  {journal} {Astrophys. J.}\ }\textbf {\bibinfo {volume} {431}},\
  \bibinfo {pages} {495} (\bibinfo {year} {1994})},\ \Eprint
  {http://arxiv.org/abs/astro-ph/9311070} {arXiv:astro-ph/9311070 [astro-ph]}
  \BibitemShut {NoStop}%
%%CITATION = ASTRO-PH/9311070;%%
\bibitem [{\citenamefont {{Scoccimarro}}\ and\ \citenamefont
  {{Frieman}}(1996{\natexlab{a}})}]{1996ApJS..105...37S}%
  \BibitemOpen
  \bibfield  {author} {\bibinfo {author} {\bibfnamefont {R.}~\bibnamefont
  {{Scoccimarro}}}\ and\ \bibinfo {author} {\bibfnamefont {J.}~\bibnamefont
  {{Frieman}}},\ }\href {\doibase 10.1086/192306} {\bibfield  {journal}
  {\bibinfo  {journal} {\apjs}\ }\textbf {\bibinfo {volume} {105}},\ \bibinfo
  {pages} {37} (\bibinfo {year} {1996}{\natexlab{a}})},\ \Eprint
  {http://arxiv.org/abs/astro-ph/9509047} {astro-ph/9509047} \BibitemShut
  {NoStop}%
\bibitem [{\citenamefont {{Scoccimarro}}\ and\ \citenamefont
  {{Frieman}}(1996{\natexlab{b}})}]{1996ApJ...473..620S}%
  \BibitemOpen
  \bibfield  {author} {\bibinfo {author} {\bibfnamefont {R.}~\bibnamefont
  {{Scoccimarro}}}\ and\ \bibinfo {author} {\bibfnamefont {J.~A.}\ \bibnamefont
  {{Frieman}}},\ }\href {\doibase 10.1086/178177} {\bibfield  {journal}
  {\bibinfo  {journal} {\apj}\ }\textbf {\bibinfo {volume} {473}},\ \bibinfo
  {pages} {620} (\bibinfo {year} {1996}{\natexlab{b}})},\ \Eprint
  {http://arxiv.org/abs/astro-ph/9602070} {astro-ph/9602070} \BibitemShut
  {NoStop}%
\bibitem [{\citenamefont {Blas}\ \emph {et~al.}(2013)\citenamefont {Blas},
  \citenamefont {Garny},\ and\ \citenamefont {Konstandin}}]{Blas:2013bpa}%
  \BibitemOpen
  \bibfield  {author} {\bibinfo {author} {\bibfnamefont {D.}~\bibnamefont
  {Blas}}, \bibinfo {author} {\bibfnamefont {M.}~\bibnamefont {Garny}}, \ and\
  \bibinfo {author} {\bibfnamefont {T.}~\bibnamefont {Konstandin}},\ }\href
  {\doibase 10.1088/1475-7516/2013/09/024} {\bibfield  {journal} {\bibinfo
  {journal} {JCAP}\ }\textbf {\bibinfo {volume} {1309}},\ \bibinfo {pages}
  {024} (\bibinfo {year} {2013})},\ \Eprint {http://arxiv.org/abs/1304.1546}
  {arXiv:1304.1546 [astro-ph.CO]} \BibitemShut {NoStop}%
%%CITATION = ARXIV:1304.1546;%%
\bibitem [{\citenamefont {Zeldovich}(1970)}]{Zeldovich:1969sb}%
  \BibitemOpen
  \bibfield  {author} {\bibinfo {author} {\bibfnamefont {{\relax Ya}.~B.}\
  \bibnamefont {Zeldovich}},\ }\href@noop {} {\bibfield  {journal} {\bibinfo
  {journal} {Astron. Astrophys.}\ }\textbf {\bibinfo {volume} {5}},\ \bibinfo
  {pages} {84} (\bibinfo {year} {1970})}\BibitemShut {NoStop}%
%%CITATION = AAEJA,5,84;%%
\bibitem [{\citenamefont {Bouchet}\ \emph {et~al.}(1995)\citenamefont
  {Bouchet}, \citenamefont {Colombi}, \citenamefont {Hivon},\ and\
  \citenamefont {Juszkiewicz}}]{Bouchet:1994xp}%
  \BibitemOpen
  \bibfield  {author} {\bibinfo {author} {\bibfnamefont {F.~R.}\ \bibnamefont
  {Bouchet}}, \bibinfo {author} {\bibfnamefont {S.}~\bibnamefont {Colombi}},
  \bibinfo {author} {\bibfnamefont {E.}~\bibnamefont {Hivon}}, \ and\ \bibinfo
  {author} {\bibfnamefont {R.}~\bibnamefont {Juszkiewicz}},\ }\href@noop {}
  {\bibfield  {journal} {\bibinfo  {journal} {Astron. Astrophys.}\ }\textbf
  {\bibinfo {volume} {296}},\ \bibinfo {pages} {575} (\bibinfo {year}
  {1995})},\ \Eprint {http://arxiv.org/abs/astro-ph/9406013}
  {arXiv:astro-ph/9406013 [astro-ph]} \BibitemShut {NoStop}%
%%CITATION = ASTRO-PH/9406013;%%
\bibitem [{\citenamefont {Matsubara}(2008)}]{Matsubara:2007wj}%
  \BibitemOpen
  \bibfield  {author} {\bibinfo {author} {\bibfnamefont {T.}~\bibnamefont
  {Matsubara}},\ }\href {\doibase 10.1103/PhysRevD.77.063530} {\bibfield
  {journal} {\bibinfo  {journal} {Phys. Rev.}\ }\textbf {\bibinfo {volume}
  {D77}},\ \bibinfo {pages} {063530} (\bibinfo {year} {2008})},\ \Eprint
  {http://arxiv.org/abs/0711.2521} {arXiv:0711.2521 [astro-ph]} \BibitemShut
  {NoStop}%
%%CITATION = ARXIV:0711.2521;%%
\bibitem [{\citenamefont {{Bernardeau}}\ \emph {et~al.}(2002)\citenamefont
  {{Bernardeau}}, \citenamefont {{Colombi}}, \citenamefont {{Gazta{\~n}aga}},\
  and\ \citenamefont {{Scoccimarro}}}]{bernardeauReview}%
  \BibitemOpen
  \bibfield  {author} {\bibinfo {author} {\bibfnamefont {F.}~\bibnamefont
  {{Bernardeau}}}, \bibinfo {author} {\bibfnamefont {S.}~\bibnamefont
  {{Colombi}}}, \bibinfo {author} {\bibfnamefont {E.}~\bibnamefont
  {{Gazta{\~n}aga}}}, \ and\ \bibinfo {author} {\bibfnamefont {R.}~\bibnamefont
  {{Scoccimarro}}},\ }\href {\doibase 10.1016/S0370-1573(02)00135-7} {\bibfield
   {journal} {\bibinfo  {journal} {Physics reports}\ }\textbf {\bibinfo
  {volume} {367}},\ \bibinfo {pages} {1} (\bibinfo {year} {2002})},\ \Eprint
  {http://arxiv.org/abs/astro-ph/0112551} {astro-ph/0112551} \BibitemShut
  {NoStop}%
\bibitem [{\citenamefont {{Carlson}}\ \emph {et~al.}(2009)\citenamefont
  {{Carlson}}, \citenamefont {{White}},\ and\ \citenamefont
  {{Padmanabhan}}}]{CarlsonWhitePadmanabhan0905}%
  \BibitemOpen
  \bibfield  {author} {\bibinfo {author} {\bibfnamefont {J.}~\bibnamefont
  {{Carlson}}}, \bibinfo {author} {\bibfnamefont {M.}~\bibnamefont {{White}}},
  \ and\ \bibinfo {author} {\bibfnamefont {N.}~\bibnamefont {{Padmanabhan}}},\
  }\href {\doibase 10.1103/PhysRevD.80.043531} {\bibfield  {journal} {\bibinfo
  {journal} {\prd}\ }\textbf {\bibinfo {volume} {80}},\ \bibinfo {eid} {043531}
  (\bibinfo {year} {2009})},\ \Eprint {http://arxiv.org/abs/0905.0479}
  {arXiv:0905.0479 [astro-ph.CO]} \BibitemShut {NoStop}%
\bibitem [{\citenamefont {{Seljak}}\ and\ \citenamefont
  {{Zaldarriaga}}(1996)}]{cmbfast}%
  \BibitemOpen
  \bibfield  {author} {\bibinfo {author} {\bibfnamefont {U.}~\bibnamefont
  {{Seljak}}}\ and\ \bibinfo {author} {\bibfnamefont {M.}~\bibnamefont
  {{Zaldarriaga}}},\ }\href {\doibase 10.1086/177793} {\bibfield  {journal}
  {\bibinfo  {journal} {\apj}\ }\textbf {\bibinfo {volume} {469}},\ \bibinfo
  {pages} {437} (\bibinfo {year} {1996})},\ \Eprint
  {http://arxiv.org/abs/astro-ph/9603033} {astro-ph/9603033} \BibitemShut
  {NoStop}%
\bibitem [{\citenamefont {Lewis}\ \emph {et~al.}(2000)\citenamefont {Lewis},
  \citenamefont {Challinor},\ and\ \citenamefont {Lasenby}}]{camb}%
  \BibitemOpen
  \bibfield  {author} {\bibinfo {author} {\bibfnamefont {A.}~\bibnamefont
  {Lewis}}, \bibinfo {author} {\bibfnamefont {A.}~\bibnamefont {Challinor}}, \
  and\ \bibinfo {author} {\bibfnamefont {A.}~\bibnamefont {Lasenby}},\
  }\href@noop {} {\bibfield  {journal} {\bibinfo  {journal} {Astrophys. J.}\
  }\textbf {\bibinfo {volume} {538}},\ \bibinfo {pages} {473} (\bibinfo {year}
  {2000})},\ \Eprint {http://arxiv.org/abs/astro-ph/9911177} {astro-ph/9911177}
  \BibitemShut {NoStop}%
%%CITATION = ASTRO-PH 9911177;%%
\bibitem [{\citenamefont {{Fendt}}\ and\ \citenamefont
  {{Wandelt}}(2007)}]{pico}%
  \BibitemOpen
  \bibfield  {author} {\bibinfo {author} {\bibfnamefont {W.~A.}\ \bibnamefont
  {{Fendt}}}\ and\ \bibinfo {author} {\bibfnamefont {B.~D.}\ \bibnamefont
  {{Wandelt}}},\ }\href {\doibase 10.1086/508342} {\bibfield  {journal}
  {\bibinfo  {journal} {\apj}\ }\textbf {\bibinfo {volume} {654}},\ \bibinfo
  {pages} {2} (\bibinfo {year} {2007})},\ \Eprint
  {http://arxiv.org/abs/astro-ph/0606709} {astro-ph/0606709} \BibitemShut
  {NoStop}%
\bibitem [{\citenamefont {{Hamilton}}(2000)}]{hamiltonfftlog}%
  \BibitemOpen
  \bibfield  {author} {\bibinfo {author} {\bibfnamefont {A.~J.~S.}\
  \bibnamefont {{Hamilton}}},\ }\href {\doibase
  10.1046/j.1365-8711.2000.03071.x} {\bibfield  {journal} {\bibinfo  {journal}
  {\mnras}\ }\textbf {\bibinfo {volume} {312}},\ \bibinfo {pages} {257}
  (\bibinfo {year} {2000})},\ \Eprint {http://arxiv.org/abs/astro-ph/9905191}
  {astro-ph/9905191} \BibitemShut {NoStop}%
\bibitem [{\citenamefont {{Vlah}}\ \emph {et~al.}(2015)\citenamefont {{Vlah}},
  \citenamefont {{Seljak}},\ and\ \citenamefont {{Baldauf}}}]{zvonimir1410}%
  \BibitemOpen
  \bibfield  {author} {\bibinfo {author} {\bibfnamefont {Z.}~\bibnamefont
  {{Vlah}}}, \bibinfo {author} {\bibfnamefont {U.}~\bibnamefont {{Seljak}}}, \
  and\ \bibinfo {author} {\bibfnamefont {T.}~\bibnamefont {{Baldauf}}},\ }\href
  {\doibase 10.1103/PhysRevD.91.023508} {\bibfield  {journal} {\bibinfo
  {journal} {\prd}\ }\textbf {\bibinfo {volume} {91}},\ \bibinfo {eid} {023508}
  (\bibinfo {year} {2015})},\ \Eprint {http://arxiv.org/abs/1410.1617}
  {arXiv:1410.1617} \BibitemShut {NoStop}%
\bibitem [{\citenamefont {Vlah}\ \emph {et~al.}(2015)\citenamefont {Vlah},
  \citenamefont {White},\ and\ \citenamefont {Aviles}}]{Vlah:2015sea}%
  \BibitemOpen
  \bibfield  {author} {\bibinfo {author} {\bibfnamefont {Z.}~\bibnamefont
  {Vlah}}, \bibinfo {author} {\bibfnamefont {M.}~\bibnamefont {White}}, \ and\
  \bibinfo {author} {\bibfnamefont {A.}~\bibnamefont {Aviles}},\ }\href
  {\doibase 10.1088/1475-7516/2015/09/014} {\bibfield  {journal} {\bibinfo
  {journal} {JCAP}\ }\textbf {\bibinfo {volume} {1509}},\ \bibinfo {pages}
  {014} (\bibinfo {year} {2015})},\ \Eprint {http://arxiv.org/abs/1506.05264}
  {arXiv:1506.05264 [astro-ph.CO]} \BibitemShut {NoStop}%
%%CITATION = ARXIV:1506.05264;%%
\bibitem [{\citenamefont {{Sherwin}}\ and\ \citenamefont
  {{Zaldarriaga}}(2012)}]{sherwinZaldarriaga}%
  \BibitemOpen
  \bibfield  {author} {\bibinfo {author} {\bibfnamefont {B.~D.}\ \bibnamefont
  {{Sherwin}}}\ and\ \bibinfo {author} {\bibfnamefont {M.}~\bibnamefont
  {{Zaldarriaga}}},\ }\href {\doibase 10.1103/PhysRevD.85.103523} {\bibfield
  {journal} {\bibinfo  {journal} {\prd}\ }\textbf {\bibinfo {volume} {85}},\
  \bibinfo {eid} {103523} (\bibinfo {year} {2012})},\ \Eprint
  {http://arxiv.org/abs/1202.3998} {arXiv:1202.3998 [astro-ph.CO]} \BibitemShut
  {NoStop}%
\bibitem [{\citenamefont {{McCullagh}}\ and\ \citenamefont
  {{Szalay}}(2012)}]{Mccollough1202}%
  \BibitemOpen
  \bibfield  {author} {\bibinfo {author} {\bibfnamefont {N.}~\bibnamefont
  {{McCullagh}}}\ and\ \bibinfo {author} {\bibfnamefont {A.~S.}\ \bibnamefont
  {{Szalay}}},\ }\href {\doibase 10.1088/0004-637X/752/1/21} {\bibfield
  {journal} {\bibinfo  {journal} {\apj}\ }\textbf {\bibinfo {volume} {752}},\
  \bibinfo {eid} {21} (\bibinfo {year} {2012})},\ \Eprint
  {http://arxiv.org/abs/1202.1306} {arXiv:1202.1306 [astro-ph.CO]} \BibitemShut
  {NoStop}%
\bibitem [{\citenamefont {{B{\"o}hm}}\ \emph {et~al.}(2016)\citenamefont
  {{B{\"o}hm}}, \citenamefont {{Schmittfull}},\ and\ \citenamefont
  {{Sherwin}}}]{BoehmN32}%
  \BibitemOpen
  \bibfield  {author} {\bibinfo {author} {\bibfnamefont {V.}~\bibnamefont
  {{B{\"o}hm}}}, \bibinfo {author} {\bibfnamefont {M.}~\bibnamefont
  {{Schmittfull}}}, \ and\ \bibinfo {author} {\bibfnamefont {B.~D.}\
  \bibnamefont {{Sherwin}}},\ }\href@noop {} {\bibfield  {journal} {\bibinfo
  {journal} {ArXiv e-prints}\ } (\bibinfo {year} {2016})},\ \Eprint
  {http://arxiv.org/abs/1605.01392} {arXiv:1605.01392} \BibitemShut {NoStop}%
\bibitem [{\citenamefont {{Fergusson}}\ \emph {et~al.}(2010)\citenamefont
  {{Fergusson}}, \citenamefont {{Liguori}},\ and\ \citenamefont
  {{Shellard}}}]{fergusson0912}%
  \BibitemOpen
  \bibfield  {author} {\bibinfo {author} {\bibfnamefont {J.~R.}\ \bibnamefont
  {{Fergusson}}}, \bibinfo {author} {\bibfnamefont {M.}~\bibnamefont
  {{Liguori}}}, \ and\ \bibinfo {author} {\bibfnamefont {E.~P.~S.}\
  \bibnamefont {{Shellard}}},\ }\href {\doibase 10.1103/PhysRevD.82.023502}
  {\bibfield  {journal} {\bibinfo  {journal} {\prd}\ }\textbf {\bibinfo
  {volume} {82}},\ \bibinfo {eid} {023502} (\bibinfo {year} {2010})},\ \Eprint
  {http://arxiv.org/abs/0912.5516} {arXiv:0912.5516} \BibitemShut {NoStop}%
\bibitem [{\citenamefont {{Fergusson}}\ \emph {et~al.}(2012)\citenamefont
  {{Fergusson}}, \citenamefont {{Regan}},\ and\ \citenamefont
  {{Shellard}}}]{fergusson1008}%
  \BibitemOpen
  \bibfield  {author} {\bibinfo {author} {\bibfnamefont {J.~R.}\ \bibnamefont
  {{Fergusson}}}, \bibinfo {author} {\bibfnamefont {D.~M.}\ \bibnamefont
  {{Regan}}}, \ and\ \bibinfo {author} {\bibfnamefont {E.~P.~S.}\ \bibnamefont
  {{Shellard}}},\ }\href {\doibase 10.1103/PhysRevD.86.063511} {\bibfield
  {journal} {\bibinfo  {journal} {\prd}\ }\textbf {\bibinfo {volume} {86}},\
  \bibinfo {eid} {063511} (\bibinfo {year} {2012})},\ \Eprint
  {http://arxiv.org/abs/1008.1730} {arXiv:1008.1730 [astro-ph.CO]} \BibitemShut
  {NoStop}%
\bibitem [{\citenamefont {{Slepian}}\ and\ \citenamefont
  {{Eisenstein}}(2015)}]{slepian1411}%
  \BibitemOpen
  \bibfield  {author} {\bibinfo {author} {\bibfnamefont {Z.}~\bibnamefont
  {{Slepian}}}\ and\ \bibinfo {author} {\bibfnamefont {D.~J.}\ \bibnamefont
  {{Eisenstein}}},\ }\href {\doibase 10.1093/mnras/stu2627} {\bibfield
  {journal} {\bibinfo  {journal} {\mnras}\ }\textbf {\bibinfo {volume} {448}},\
  \bibinfo {pages} {9} (\bibinfo {year} {2015})},\ \Eprint
  {http://arxiv.org/abs/1411.4052} {arXiv:1411.4052} \BibitemShut {NoStop}%
\bibitem [{\citenamefont {{Smith}}\ \emph {et~al.}(2015)\citenamefont
  {{Smith}}, \citenamefont {{Senatore}},\ and\ \citenamefont
  {{Zaldarriaga}}}]{kendrickTrisp1502}%
  \BibitemOpen
  \bibfield  {author} {\bibinfo {author} {\bibfnamefont {K.~M.}\ \bibnamefont
  {{Smith}}}, \bibinfo {author} {\bibfnamefont {L.}~\bibnamefont {{Senatore}}},
  \ and\ \bibinfo {author} {\bibfnamefont {M.}~\bibnamefont {{Zaldarriaga}}},\
  }\href@noop {} {\bibfield  {journal} {\bibinfo  {journal} {ArXiv e-prints}\ }
  (\bibinfo {year} {2015})},\ \Eprint {http://arxiv.org/abs/1502.00635}
  {arXiv:1502.00635} \BibitemShut {NoStop}%
\bibitem [{\citenamefont {{Carrasco}}\ \emph {et~al.}(2014)\citenamefont
  {{Carrasco}}, \citenamefont {{Foreman}}, \citenamefont {{Green}},\ and\
  \citenamefont {{Senatore}}}]{Carrasco1304}%
  \BibitemOpen
  \bibfield  {author} {\bibinfo {author} {\bibfnamefont {J.~J.~M.}\
  \bibnamefont {{Carrasco}}}, \bibinfo {author} {\bibfnamefont
  {S.}~\bibnamefont {{Foreman}}}, \bibinfo {author} {\bibfnamefont
  {D.}~\bibnamefont {{Green}}}, \ and\ \bibinfo {author} {\bibfnamefont
  {L.}~\bibnamefont {{Senatore}}},\ }\href {\doibase
  10.1088/1475-7516/2014/07/056} {\bibfield  {journal} {\bibinfo  {journal}
  {\jcap}\ }\textbf {\bibinfo {volume} {7}},\ \bibinfo {eid} {056} (\bibinfo
  {year} {2014})},\ \Eprint {http://arxiv.org/abs/1304.4946} {arXiv:1304.4946}
  \BibitemShut {NoStop}%
\bibitem [{NIS()}]{NISTDLMF}%
  \BibitemOpen
  \href {http://dlmf.nist.gov/} {\enquote {\bibinfo {title} {{NIST Digital
  Library of Mathematical Functions}},}\ }\bibinfo {howpublished}
  {http://dlmf.nist.gov/, Release 1.0.5 of 2012-10-01},\ \bibinfo {note}
  {online companion to \cite{Olver:2010:NHMF}}\BibitemShut {NoStop}%
\bibitem [{\citenamefont {Olver}\ \emph {et~al.}(2010)\citenamefont {Olver},
  \citenamefont {Lozier}, \citenamefont {Boisvert},\ and\ \citenamefont
  {Clark}}]{Olver:2010:NHMF}%
  \BibitemOpen
  \bibinfo {editor} {\bibfnamefont {F.~W.~J.}\ \bibnamefont {Olver}}, \bibinfo
  {editor} {\bibfnamefont {D.~W.}\ \bibnamefont {Lozier}}, \bibinfo {editor}
  {\bibfnamefont {R.~F.}\ \bibnamefont {Boisvert}}, \ and\ \bibinfo {editor}
  {\bibfnamefont {C.~W.}\ \bibnamefont {Clark}},\ eds.,\ \href@noop {} {\emph
  {\bibinfo {title} {{NIST Handbook of Mathematical Functions}}}}\ (\bibinfo
  {publisher} {Cambridge University Press},\ \bibinfo {address} {New York,
  NY},\ \bibinfo {year} {2010})\ \bibinfo {note} {print companion to
  \cite{NISTDLMF}}\BibitemShut {NoStop}%
\bibitem [{\citenamefont {Watson}(1966)}]{Watson66}%
  \BibitemOpen
  \bibfield  {author} {\bibinfo {author} {\bibfnamefont {G.~N.}\ \bibnamefont
  {Watson}},\ }\href@noop {} {\emph {\bibinfo {title} {{A Treatise on the
  Theory of Bessel Functions}}}}\ (\bibinfo  {publisher} {Cambridge University
  Press},\ \bibinfo {year} {1966})\BibitemShut {NoStop}%
\bibitem [{\citenamefont {{Matsubara}}(2008)}]{Matsubara0807}%
  \BibitemOpen
  \bibfield  {author} {\bibinfo {author} {\bibfnamefont {T.}~\bibnamefont
  {{Matsubara}}},\ }\href {\doibase 10.1103/PhysRevD.78.083519} {\bibfield
  {journal} {\bibinfo  {journal} {\prd}\ }\textbf {\bibinfo {volume} {78}},\
  \bibinfo {eid} {083519} (\bibinfo {year} {2008})},\ \Eprint
  {http://arxiv.org/abs/0807.1733} {arXiv:0807.1733} \BibitemShut {NoStop}%
\end{thebibliography}%

\end{document}